\documentclass[aps,prd,reprint,superscriptaddress,nofootinbib,floatfix,longbibliography]{revtex4-1}
\usepackage[dvipsnames]{xcolor}
\usepackage{amssymb,amsmath,amsfonts,bm,slashed,soul,ragged2e,graphicx,epstopdf,hyperref,array,bm}
\usepackage[utf8]{inputenc}
\usepackage{colortbl,color,caption,subcaption}

\enlargethispage{\baselineskip}
\definecolor{steelblue}{RGB}{25,25,112}
\definecolor{dullblue}{rgb}{0,0.298,0.49}
\definecolor{darkred}{rgb}{0.545,0,0}
\definecolor{darkorange}{RGB}{222,132,69}
\definecolor{darkgreen}{RGB}{126,171,85}
\definecolor{blue2}{cmyk}{1, 0.1, 0.1, 0}
\hypersetup{colorlinks,linkcolor={darkred},citecolor={blue},urlcolor={dullblue}}
\DeclareCaptionJustification{justified}{\justifying}
\everymath{\displaystyle}
\usepackage{placeins}

\usepackage{comment}
\usepackage{braket}
\usepackage{amsmath}
\usepackage{soul}
\usepackage{makecell}
\usepackage{booktabs}
\usepackage{siunitx}
\usepackage{multirow}
\usepackage{cancel}

\newcommand{\beq}{\begin{equation}}
\newcommand{\eeq}{\end{equation}}
\newcommand{\bea}{\begin{eqnarray}}
\newcommand{\eea}{\end{eqnarray}}

\newcommand{\gsim}{\lower.7ex\hbox{$\;\stackrel{\textstyle>}{\sim}\;$}}
\newcommand{\lsim}{\lower.7ex\hbox{$\;\stackrel{\textstyle<}{\sim}\;$}}

\newcommand{\be}{\begin{equation}}
\newcommand{\ee}{\end{equation}}
\newcommand{\ba}{\begin{eqnarray}}
\newcommand{\ea}{\end{eqnarray}}

\def\d{\mathrm{d}}

\def\A{\mathcal{A}}

\def\vec{\mathbf}

\def\P{\mathcal{P}}

\def\Ainf{A_{s}^{\rm inf}}
\def\Acurv{A_{s}^{\rm curv}}
\def\evphi{\langle \phi \rangle}

\def\vaphi{\langle \phi^2 \rangle}
\def\evvphi{\langle \varphi \rangle}
\def\vavphi{\langle \varphi^2 \rangle}
\def\vasigma{\langle \sigma^2 \rangle}
\newcommand{\Cubic}{{\it cubic} }
\newcommand{\Quartic}{{\it quartic} }

\RequirePackage[normalem]{ulem}

\usepackage{orcidlink}

\begin{document}

\title{Curvaton-assisted hilltop inflation}

\author{Wen-Yuan Ai\,\orcidlink{0000-0002-6042-7407}}
 \email{wenyuanai@sjtu.edu.cn}
\affiliation{
State Key Laboratory of Dark Matter Physics,\\ Tsung-Dao Lee Institute and School of Physics and Astronomy,\\ Shanghai Jiao Tong University, Shanghai 201210, China}

\affiliation{Key Laboratory for Particle Astrophysics and Cosmology (MOE),\\ and Shanghai Key Laboratory for Particle Physics and Cosmology,\\ Shanghai Jiao Tong University, Shanghai 201210, China}

\author{Stephen F. King\,\orcidlink{0000-0002-4351-7507}}
\email{king@soton.ac.uk}
\affiliation{School of Physics and Astronomy, University of Southampton, Southampton SO17 1BJ, United Kingdom}

\author{Xin Wang\,\orcidlink{0000-0003-4292-460X}}
\email{xin.wang@unipd.it}
\affiliation{School of Physics and Astronomy, University of Southampton, Southampton SO17 1BJ, United Kingdom}
\affiliation{Dipartimento di Fisica e Astronomia ``Galileo Galilei'', Universit\`a degli Studi di Padova,
Via Francesco Marzolo 8, 35131 Padova, Italy}
\affiliation{INFN, Sezione di Padova, Via Francesco Marzolo 8, 35131 Padova, Italy}

\author{Ye-Ling Zhou\,\orcidlink{0000-0002-3664-9472}}
\email{zhouyeling@ucas.ac.cn}
\affiliation{School of Fundamental Physics and Mathematical Sciences, Hangzhou Institute for Advanced Study, UCAS,
Hangzhou 310024, China}

\begin{abstract}
Following the recent Atacama Cosmology Telescope (ACT) results, we consider hilltop inflation where the inflaton is coupled to a curvaton, simultaneously addressing two main challenges faced by conventional hilltop inflation models: the initial-value problem; and their viability for sub-Planckian field values. In standard single-field hilltop inflation, the inflaton must start extremely close to the maximum of the potential, raising concerns about the naturalness of the initial conditions. We demonstrate that the curvaton field not only significantly relaxes the initial-value tuning required for hilltop inflation, but also opens up parameter space through modifying the curvature perturbation power spectrum, reviving  the {\it quartic} hilltop inflation model in the sub-Planckian regime. We find viable parameter space consistent with the recent cosmological observations.

\end{abstract}

\date{\today}

\maketitle

\section{Introduction}
Precision measurements of primordial curvature perturbations have established a spectrum that is almost scale-invariant, predominantly adiabatic and close to Gaussian, with a small red tilt~\cite{Planck:2013jfk,Planck:2018jri,Planck:2019kim}. This strongly favors the inflationary paradigm, in which a scalar field (the inflaton) slowly rolls along a very flat region of its potential~\cite{Starobinsky:1980te,Guth:1980zm,Linde:1981mu,Albrecht:1982wi}. 

Recently, the most up-to-date Atacama Cosmology Telescope (ACT) DR6 analysis has reported a spectral index of $n_s=0.9743\pm 0.0034$ based on a joint fit with Planck and the Dark Energy Spectroscopic Instrument (DESI) DR1 data~\cite{ACT:2025fju}. This value differs from the original \texttt{Planck18} (TT,TE,EE+lowE) result $n_s=0.9649\pm 0.0044$~\cite{Planck:2018jri} by about $2\sigma$ and has triggered a lot of discussions on inflationary models, see, e.g., Refs.~\cite{Kallosh:2025rni,Aoki:2025wld,Berera:2025vsu,Dioguardi:2025vci,Gialamas:2025kef,Salvio:2025izr,Dioguardi:2025mpp,He:2025bli,Drees:2025ngb,Kim:2025dyi,Gialamas:2025ofz,Antoniadis:2025pfa,Liu:2025qca,Zharov:2025evb,Yogesh:2025wak,Addazi:2025qra,Yin:2025rrs,Byrnes:2025kit,Han:2025cwk,Mohammadi:2025gbu,Lynker:2025wyc,Keus:2025iwa}.

In the context of inflationary model building, hilltop inflation~\cite{Linde:1981mu,Albrecht:1982wi,Izawa:1996dv,Senoguz:2004ky,Boubekeur:2005zm} constitutes an attractive class of models, in which inflation is driven by a scalar field $\phi$ rolling down from the vicinity of an unstable local maximum of the potential, which can be parametrized near the hilltop as
\begin{align}
    V(\phi)=\Lambda^4(1-\phi^p/\mu^p + \cdots) \; ,
    \label{eq:hiltop-general}
\end{align}
where $\Lambda$ and $\mu$ are characteristic energy scales, and $p \geq 2$ denotes an integer. Recently, it has been shown that hilltop inflation can be realized in a modular-invariant framework if the inflaton is identified with a modulus field~\cite{King:2024ssx, Ding:2024neh}.
The {\it quadratic} case ($p=2$) is viable only as a large-field inflation model with $\mu \gtrsim M_{\rm Pl}$. By contrast, for the \Cubic ($p=3$) and \Quartic ($p=4$) hilltop potentials, the entire sub-Planckian regime ($\mu \lesssim M_{\rm Pl}$) already lies outside the 95\% confidence level (C.L.) region of the Planck constraints~\cite{Planck:2018jri}.

On the other hand, since the inflation is required to start extremely close to the hilltop of the potential, hilltop inflation clearly faces an initial condition problem. In Ref.~\cite{Antusch:2014qqa}, this problem was addressed by introducing a pre-inflationary stage during which a matter field undergoes classical slow-roll evolution and dynamically drags the inflaton toward the suitable initial position for the subsequent hilltop inflationary phase. 

In this work, motivated by these problems, we consider a curvaton-assisted hilltop scenario in which the inflaton is coupled to a light scalar field $\sigma$, whose dynamics in the pre-inflationary era is dominated by quantum diffusion. We track the evolution of $\sigma$ and $\phi$ during the pre-inflationary stage by solving the Langevin equations in the quantum-diffusion region, and derive analytically a characteristic onset window for hilltop inflation that is only weakly sensitive to the initial value of $\sigma$. Once hilltop inflation starts, $\sigma$ becomes effectively frozen and plays the role of a curvaton~\cite{Linde:1996gt, Lyth:2001nq,  Moroi:2001ct, Moroi:2002rd, Lyth:2002my}, modifying curvature perturbations upon the curvaton decay at later times after inflation and alleviating the tension of conventional hilltop models with current Cosmic Microwave  Background (CMB) observations.
We perform an illustrative Bayesian analysis based on a simplified likelihood constructed from the Planck and ACT results, assessing and comparing how well our model is supported by each of them. From the resulting posterior distributions, we further present the model predictions for the tensor-to-scalar ratio and primordial non-Gaussianity. We demonstrate that the curvaton field not only significantly relaxes the initial-value tuning of hilltop inflation, but also opens up parameter space through modifying the curvature perturbation power spectrum, reviving the {\it quartic} hilltop inflation models in the sub-Planckian regime.

The layout of the remainder of this paper is as follows. We construct our model in Sec.~\ref{sec:model}. In Sec.~\ref{sec:initial-condition}, we derive the initial conditions for hilltop inflation. In Sec.~\ref{sec:curvaton}, we discuss how the curvaton mechanism modifies the primordial observables in hilltop inflation. In Sec.~\ref{sec:numerical-results}, we show the results of our Bayesian analysis and discuss their implications. We summarize our main conclusions in Sec.~\ref{sec:con}. In appendices~\ref{app:langevin} and \ref{app:deltaN}, we provide additional technical details of Langevin equations and the $\delta N$ formalism. In appendix~\ref{app:cubic}, we discuss the viability of the \Cubic hilltop case.

\section{Hilltop inflation with a curvaton}\label{sec:model}
We consider a two-field scalar potential involving an inflaton $\phi$ and a curvaton $\sigma$. $\phi$ drives hilltop inflation during the slow-roll phase, whereas $\sigma$ modifies the power spectrum of primordial perturbations after inflation. The scalar potential can be expressed in a general form as
\begin{align}
    V = V_{\rm HT} + V_{\rm S} + V_{\rm C}  \; ,
    \label{eq:potential}
\end{align}
where $V_{\rm HT}$ denotes the primary hilltop potential~\cite{Linde:1981mu,Albrecht:1982wi} 
\begin{align}
    V_{\rm HT} = \Lambda^4 \left(1 -  \frac{\phi^p}{2\mu^p}  \right)^2 \; ,
    \label{eq:V_HT}
\end{align} 
where a factor of two is added in the denominator to be consistent with eq.~\eqref{eq:hiltop-general}. In this paper, we focus on two specific examples: $p=3$ (\Cubic hilltop inflation) and $p=4$ (\Quartic hilltop inflation). 

The soft term in eq.~\eqref{eq:potential} is given by
\begin{align}
     V^{}_{\rm S} = -  \frac{1}{2}m^2_\phi \phi^2_{} + \frac{1}{2} m^2_\sigma \sigma^2_{} \; ,
    \label{eq:V_M}
\end{align}
with $m_\phi$ and $m_\sigma$ being the bare masses of $\phi$ and $\sigma$, respectively. Note that $m^2_\sigma$ should be much smaller than the Hubble parameter $H^2 \approx V/(3M_{\rm Pl}^2)$ during inflation to ensure that the curvaton $\sigma$ is effectively frozen.

Finally, we use $V_{\rm C}$ to represent the cross term between $\phi$ and $\sigma$, which is chosen to be   
\begin{align}
    V^{}_{\rm C} = \frac{\lambda^2}{2 M_{\rm Pl}^2} \phi^2 \sigma^4  \; ,
    \label{eq:V_C}
\end{align}
where $\lambda$ is a dimensionless coupling constant and $M_{\rm Pl} \equiv 2.435 \times 10^{18}\,{\rm GeV}$ is the reduced Planck mass. $V_{\rm C}$ contributes to the effective mass of $\phi$ and modulates the shape of the potential along the $\phi$-direction. To be more specific, we assume that $\sigma$ starts from a sufficiently large value, so that the effective mass $m^{2}_{\phi,{\rm eff}} =  \lambda^2_{}\sigma^4_{}/M^2_{\rm Pl} - m^2_\phi > 0$ and $\phi = 0$ is a minimum in the $\phi$-direction. As $\sigma$ moves close to the critical value $\sigma_{\rm c} = (m_\phi M_{\rm Pl}/\lambda)^{1/2}$, $\phi = 0$ transitions from a local minimum to a saddle point, which naturally sets the stage for the subsequent hilltop inflation. 

Before going to the next section, we shall mention that the scalar potential discussed above could originate from supersymmetry. To be more specific, $V_{\rm HT}$ and $V_{\rm C}$ can be derived from the following superpotential
 \begin{align}
     W = \widehat{S}_1\left( \frac{\widehat{\Phi}^{p}}{M^{p-2}}-\Lambda^2\right) + \frac{2\lambda}{M_{\rm Pl}} \widehat{S}_2\widehat{\Phi}\widehat{X}^2 \; ,
 \end{align}
 where $\widehat{S}_1$, $\widehat{S}_2$, $\widehat{\Phi}$ and $\widehat{X}$ are the superfields, and $M$ is a mass parameter. These superpotential terms are enforced by arranging $S_1$ and $S_2$ to have $R$-charge 2 and imposing an additional $Z_p$ symmetry with suitable charges for $\widehat{S}_2$, $\widehat{\Phi}$ and $\widehat{X}$. It is straightforward to identify that $V_{\rm HT} = |\partial W/\partial \widehat{S}_1|^2$ and $V_{\rm C} = |\partial W/\partial \widehat{S}_2|^2$ given the relations $\phi = \sqrt{2}{\rm Re}\,\Phi$,  $\sigma = \sqrt{2}|X|$ (with $\Phi$ and $X$ being the scalar components of $\widehat{\Phi}$ and $\widehat{X}$, respectively) and $\mu^p = 2^{p/2-1}M^{p-2}\Lambda^2$, whereas $V_{\rm S}$ is attributed to soft supersymmetry breaking corrections.

\section{Initial conditions for hilltop inflation}
\label{sec:initial-condition}

In the vicinity of $\phi = 0$, the evolution of $\phi$ is dominated by the quantum fluctuations, giving rise to a quantum diffusion region whose width depends on the value of $\sigma$. Within one Hubble time, the typical displacement of $\phi$ induced by quantum diffusion is given by $\delta \phi_{\rm q} \simeq H/(2\pi)$, while the displacement due to the classical drift can be estimated as $\delta \phi_{\rm cl} \simeq |V_\phi|/(3H^2)$ with $V_\phi \equiv \partial V/\partial \phi$. The boundary $\phi_{\rm b}$ of the diffusion region can be roughly determined by equating $\delta \phi_{\rm q}$ and $\delta \phi_{\rm cl}$, namely,
\begin{align}
    \frac{H}{2\pi} = M^2_{\rm Pl} \left|\frac{V_\phi}{V}\right|_{\phi_{\rm b}} \; ,
    \label{eq:boundary}
\end{align}
where 
\begin{align}
    V^{}_\phi = m^2_\phi \left(\frac{\sigma^4_{}}{\sigma^4_{\rm c}} - 1\right) \phi - p \Lambda^4_{} \frac{\phi^{p-1}_{}}{\mu^{p}_{}} + \cdots \; , 
    \label{eq:V_phi}
\end{align}
and ``$\cdots$'' denotes higher-order corrections. We identify the onset of hilltop inflation with the exit of $\phi$ from the diffusion region. The first step is therefore to determine the value of $\sigma$ at which $\phi$ leaves the diffusion region and its dynamics becomes classical.

The evolution of $\phi$ and $\sigma$ inside the diffusion region is described by the following Langevin equations~\cite{Starobinsky:1994bd, Enqvist:2012xn, Hardwick:2017fjo}
\begin{align}
    \frac{d\phi}{dN} &= -\frac{V_\phi}{3H^2} + \frac{H}{2\pi} \xi_{\phi}(N) \; ,\label{eq:langevin_phi} \\
    \frac{d\sigma}{dN} &= -\frac{V_\sigma}{3H^2} + \frac{H}{2\pi} \xi_{\sigma}(N) \; , \label{eq:langevin_sigma}
\end{align}
where $V_\sigma \equiv \partial V/\partial \sigma$, and we use the number of e-folds $N$ as the time variable. The first terms on the right-hand sides represent classical drift, and the second terms are the stochastic ``kicks'' from quantum diffusion with $\xi_{\phi}(N)$ and $\xi_{\sigma}(N)$ being independent Gaussian white noise terms with zero mean and unit variance, i.e., $\langle \xi_i(N) \xi_j(N') \rangle = \delta_{ij}\delta(N-N')$. It should be mentioned that, unlike Ref.~\cite{Antusch:2014qqa} where the pre-inflationary direction is taken as a slow-roll direction dominated by classical drift, the curvaton direction of interest can be dominated by quantum diffusion since we require $m_\sigma \ll H$. In this case, quantum fluctuations may occasionally kick 
$\sigma$ back to larger field values, so that in certain Hubble patches inflation could keep taking place, leading to the paradigm of eternal inflation~\cite{Linde:1986fd}. In the present work, we treat eternal inflation as an open question, noting that such behavior is in fact already a generic feature of single-field hilltop inflation models~\cite{Rudelius:2019cfh}.

\begin{figure}[t!]
        \centering
        \includegraphics[width=1\linewidth]{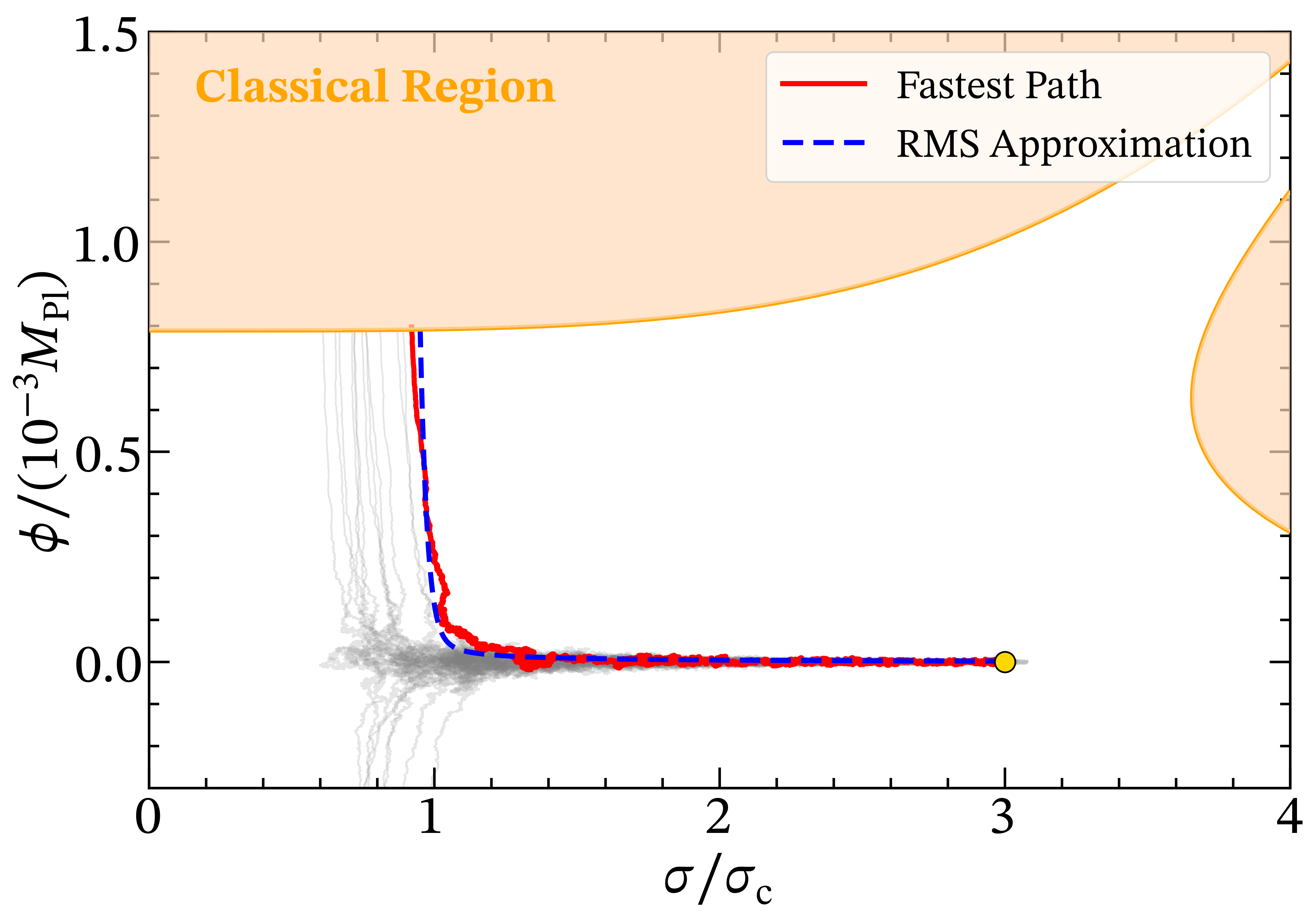}
        \caption{Comparison between the stochastic paths of fields using the Markov-chain Monte Carlo (MCMC) simulation and the root-mean-square (RMS) approximation in the \Quartic hilltop model, where we set $\Lambda = 3\times10^{-4}\,M_{\rm Pl}$, $\mu = 0.7\,M_{\rm Pl}$, $\lambda = 2 \times 10^{-4}$, $m_\phi = 3\times 10^{-11}M_{\rm Pl}$ and $m_\sigma = 5 \times 10^{-12}M_{\rm Pl}$.
       The gray lines depict the field trajectories initialized at the yellow point ($\phi_0=0, \sigma_0=3\sigma_{\rm c}$), and terminated once they cross into the classical region $[M^2_{\rm Pl}|V_\phi/V| > H/(2\pi)]$, indicated by the orange shading. The solid red line denotes the fastest escaped trajectory. The dashed blue curve represents the RMS-approximated path, obtained by solving the Langevin equations in terms of mean-square values of the fields.}
        \label{fig:trajectories}
\end{figure}

The Langevin equations given in eqs.~(\ref{eq:langevin_phi}) and (\ref{eq:langevin_sigma}) can be numerically solved using a Markov-chain Monte Carlo (MCMC) approach. To be specific, we start a large ensemble of trajectories from an initial point with $\phi_{0} = 0$ and $\sigma_0 \gtrsim \sigma_{\rm c}$. We track these trajectories until the first time they cross the diffusion boundary defined by eq.~(\ref{eq:boundary}), with the additional constraint that $V_\phi < 0$ to ensure the onset of hilltop inflation. Taking the \Quartic hilltop model for instance, we show the simulation results in Fig.~\ref{fig:trajectories}. All the gray trajectories presented in Fig.~\ref{fig:trajectories} start from the same point ($\phi_0=0, \sigma_0=3\sigma_c$), and terminate once they first reach the diffusion boundary satisfying eq.~\eqref{eq:boundary}. Since we are particularly interested in the paths that cross into the classical region at earlier times, we highlight the fastest escaped trajectory in red.

\begin{figure}[t!]
        \centering
        \includegraphics[width=\linewidth]{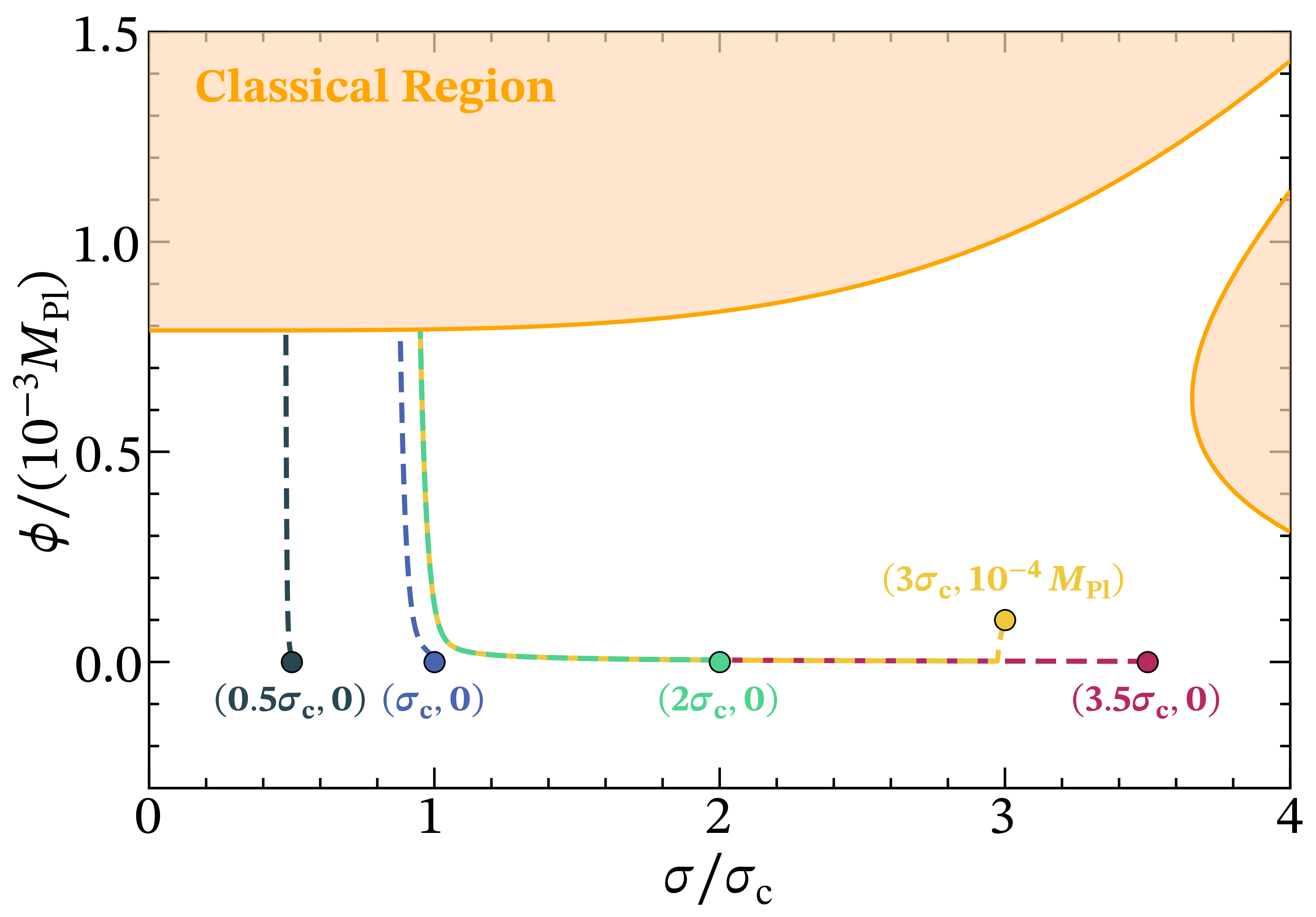}
        \caption{The RMS-approximated paths for different starting points in the \Quartic hilltop model. Model parameters in the scalar potential take the same values as those in Fig.~\ref{fig:trajectories}.}
        \label{fig:RMS-paths}
\end{figure}

By averaging Langevin equations over many realizations of the stochastic process, one can obtain equations for the mean-square values of the fields $\langle \phi^2 \rangle$ and $\langle \sigma^2 \rangle$ as
\begin{subequations}
\begin{align}
\frac{{\rm d}\langle \phi^2 \rangle}{{\rm d}N}
&= -\frac{2}{3H^2}\,\langle V_\phi \cdot\phi \rangle
+ \left(\frac{H}{2\pi}\right)^2 \; , \label{eq:langevin_phi_var}  \\
\frac{{\rm d}\langle \sigma^2 \rangle}{{\rm d} N}
&= -\frac{2}{3H^2}\,\langle V_\sigma \cdot\sigma \rangle
+ \left(\frac{H}{2\pi}\right)^2 \; .
\label{eq:langevin_sigma_var}
\end{align}
\end{subequations}
For the parameter region of interest, the covariance between $\phi$ and $\sigma$ can be safely neglected. Furthermore, given that the fields start with $\phi_0 \approx 0$ and $\sigma_0 \gtrsim \sigma_{\rm c}$, the above equations can be approximated by
\begin{subequations}
\begin{align}
    \frac{\d \vaphi }{\d N} & \approx - \frac{2}{3H^2}\left(  C_1 \vaphi + C_2 \vaphi^2  \right)+ \left(\frac{H}{2\pi}\right)^2 \; , \label{eq:langevin_phi_var_app} \\
    \frac{\d \vasigma }{\d N} & \approx - \frac{2 m^2_\sigma}{3H^2}\langle  \sigma^2 \rangle - \frac{4\lambda^2 \langle\phi^2\rangle \langle \sigma^4 \rangle}{3 H^2 M_{\rm Pl}^2}   +  \left(\frac{H}{2\pi}\right)^2 \; ,
    \label{eq:langevin_sigma_var_app}
\end{align}
\end{subequations}
where
\begin{align}
    C_1  = m^2_\phi \left(\frac{\langle\sigma^4_{}\rangle}{\sigma^4_{\rm c}}-1\right) \; ,\quad
    C_2  = -\frac{27\Lambda^8}{m^2_\phi \mu^6_{}} \left(\frac{\langle\sigma^4_{}\rangle}{\sigma^4_{\rm c}}-1\right)^{-1} ,
    \label{eq:C_cubic}
\end{align}
for $p = 3$, and
\begin{align}
    C_1  = m^2_\phi \left(\frac{\langle\sigma^4_{}\rangle}{\sigma^4_{\rm c}}-1\right) \; , \quad C_2  = -\frac{12\Lambda^4}{\mu^4_{}} \; .
    \label{eq:C_quar}
\end{align}
for $p = 4$. A detailed derivation of eqs.~\eqref{eq:langevin_phi_var}\,--\,\eqref{eq:C_quar} can be found in appendix~\ref{app:langevin}. Note that  here we take $\langle \sigma^4\rangle \simeq \langle \sigma^2 \rangle^2$, provided that $\sigma$ starts from a large value and follows an approximate Gaussian distribution.

In this formalism, the explicit noise dependence can be removed, and the competition between the diffusive source term and the classical drift term becomes manifest. We can approximately regard the positions of $\phi $ and $\sigma$ at a given time as their respective root-mean-square (RMS) values, namely, $\phi \simeq \phi_{\rm rms} = \sqrt{\langle \phi ^2 \rangle}$ and $\sigma \simeq \sigma_{\rm rms} = \sqrt{\langle \sigma ^2\rangle}$. Taking the \Quartic hilltop model as an example, we numerically solve eqs.~\eqref{eq:langevin_phi_var_app} and \eqref{eq:langevin_sigma_var_app}, and obtain an RMS-approximated path, which is denoted by the blue dashed curve in Fig.~\ref{fig:trajectories}. The RMS-approximated path agrees well with the successfully escaped paths obtained from MCMC, in particular the fastest one, indicating that it can be used to represent the stochastic behavior of the fields in the diffusion region.

We further present the RMS-approximated paths starting from different points in Fig.~\ref{fig:RMS-paths}. It is easy to see that for starting points with $\sigma_0 \lesssim  1.1\sigma_{\rm c}$, $\phi$ rapidly diffuses to the boundary, while $\sigma$ diffuses slowly. Nevertheless, if we start at some points where $\sigma_0 \gtrsim 1.1\sigma_{\rm c}$, we can clearly observe an ``L-shaped'' path: when $\sigma$ is large, $\phi$ is trapped near $\phi = 0$ by its large positive effective mass, and its evolution is strongly suppressed; once $\sigma$ approaches $1.1\sigma_{\rm c}$,  $\phi$ starts to diffuse rapidly. 
It is interesting that such transitions occur before $\sigma$ reaches $\sigma_{\rm c}$, because the second term on the right-hand side of eq.~\eqref{eq:V_phi} can provide a sizable negative contribution for large $\phi$, given the large value of $\Lambda$ in our model. In this case, even if the effective mass of $\phi$ at $\phi = 0$ remains positive, large quantum fluctuations can push the field over the barrier into the hilltop inflationary region, known as stochastic tunneling~\cite{Ellis:1990bv,Linde:1991sk,Espinosa:2007qp,Kitajima:2019ibn,Animali:2022otk}.

In fact, the ``L-shaped'' transition can be understood from the properties of Langevin equations for the field mean-square values. Before the turning point of the ``L-shaped'' trajectory, $\phi \simeq 0$. Then for the evolution of $\sigma$, the first term on the right-hand side of eq.~\eqref{eq:langevin_sigma_var_app} is dominant due to a large $\sigma_0$. As a result, we have $\sigma \approx \sigma^{}_0 \exp{(-\alpha^{}_{\sigma} N)}$ with $\alpha^{}_\sigma \equiv m^2_\sigma / (3H^2_{})$. As we focus on the case where $\alpha^{}_\sigma \ll 1$, the evolution of $\sigma$ should be slow. Meanwhile, the slow growth of $\phi$ in the regime $\sigma \gtrsim \sigma_c$ indicates the presence of a stable fixed point of eq.~\eqref{eq:langevin_phi_var_app}, where the classical drift and quantum diffusion compensate with each other. More specifically, we have
\begin{align}
    f\left( \vaphi , \sigma \right) \equiv  C_1 \vaphi + C_2 \vaphi^2 - \frac{3H^4}{8\pi^2} = 0 \; .
    \label{eq:fix_cond_1}
\end{align}
According to the bifurcation theory of differential equations, the critical condition at which the fixed point ceases to be stable is given by $\partial f(\vaphi,\sigma) / \partial \vaphi = 0$, which yields the following equation for $\vaphi$
\begin{align}
    \vaphi = -\frac{C_1}{2C_2} \; .
    \label{eq:slow_grow}
\end{align}
Substituting eq.~\eqref{eq:slow_grow} into eq.~\eqref{eq:fix_cond_1}, we arrive at the value of $\sigma_{\rm t}$ at the turning point as
\begin{align}
    \frac{\sigma_{\rm t}}{\sigma^{}_c} = \left( 1 + \frac{B_p}{m^2_\phi}\right)^{\frac{1}{4}} \; , 
    \label{eq:sigma_end}
\end{align}
with
\begin{align}
    B_3  = \left(\frac{9\Lambda^{16}_{}}{2\pi^2\mu^6_{} M^{4}_{\rm Pl}} \right)^{\frac{1}{3}}\; ,  \quad
    B_4  = \frac{\sqrt{2} \Lambda^6_{}}{\pi \mu^2 M^2_{\rm Pl}}  \; .
\end{align}

In \Quartic hilltop inflation, with the help of eq.~\eqref{eq:sigma_end}, we find that for the parameter values used in Fig.~\ref{fig:RMS-paths}, the transition emerges at  $\sigma_{\rm t} \approx 1.1\sigma_{\rm c}$, in excellent agreement with the numerical result. A similar consistency can also be observed in the case of \Cubic hilltop model. 

After the turning point, $\phi$ grows rapidly, while the term proportional to $-\langle \phi^2 \rangle \langle \sigma^4 \rangle$ in eq.~\eqref{eq:langevin_sigma_var_app} further drives $\sigma$ downward. As a result, the value of $\sigma$ at the exit from the diffusion region is slightly smaller than $\sigma_{\rm t}$.

Eq.~\eqref{eq:sigma_end} predicts a characteristic value for the exit of $\phi$ from the diffusion region that is only weakly sensitive to the initial field values, apart from the unavoidable stochasticity induced by quantum fluctuations, provided that $\sigma_0 \gtrsim \sigma_{\rm t}$ and $\phi_0 \approx 0$. Different pre-inflationary histories are driven into a relatively narrow onset window for hilltop inflation, which makes the corresponding initial condition considerably less fine-tuned. Moreover, as indicated by the numerical Langevin evolution, the actual value of $\sigma$ when the system exits the diffusion region is typically slightly smaller than the estimate $\sigma_{\rm t}$ obtained from eq.~\eqref{eq:sigma_end}. We therefore parametrize the effectively frozen curvaton field value during the subsequent hilltop inflationary stage as $\sigma_{\rm inf} \simeq \xi\,\sigma_{\rm t}$, with $\xi$ being an $\mathcal{O}(1)$ factor slightly smaller than unity. Throughout the analysis we set $\xi=0.75$, as motivated by our numerical results for the Langevin trajectories. We have further checked that the fit to the inflationary observables is not sensitive to small changes in $\xi$.

\section{Predictions for inflationary observables}
\label{sec:curvaton}

\subsection{Single-field slow-roll inflation}

In the inflationary picture, quantum vacuum fluctuations on sub-horizon scales induce small anisotropies in the otherwise homogeneous background. During inflation, the comoving Hubble radius $\sim 1/(aH)$ (with $a$ being the scale factor) decreases and eventually becomes smaller than the comoving wavelength of a given mode. As a result, the quantum
fluctuations are stretched to super-horizon scales and become effectively frozen until the end of inflation. For a mode with comoving wave number $k$, its perturbations after horizon exit can be described by a classical probability distribution, whose statistical properties are determined by the power spectrum evaluated at horizon crossing $k = aH$. In the single-field inflationary scenario, the power spectrum for scalar perturbations can be expressed as
\begin{align}
    {\cal P}_{\zeta}^{\rm inf} = \left.\frac{H^2}{4\pi^2} \frac{H^2}{\dot{\phi}^2}\right|_{k=aH} \; ,
    \label{eq:PS_scalar}
\end{align}
which relates to the primordial curvature perturbations. For tensor perturbations, we have
\begin{align}
    {\cal P}_{\cal T}^{\rm inf} = \left.\frac{2}{\pi^2_{}} \frac{H^2}{M^2_{\rm Pl}}\right|_{k=aH} \; .
    \label{eq:PS_tensor}
\end{align}

By convention, the primordial power spectra are usually parametrized as power laws, namely,
\begin{subequations}
\begin{align}
    \mathcal{P}_{\zeta}^{\rm inf}(k) &= A_s \left(\frac{k}{k_*}\right)^{n_s-1 +  \cdots} \; ,    \label{eq:Ptensor_powerlaw} \\
    \mathcal{P}_{\cal T}^{\rm inf}(k) &= r\,A_s \left(\frac{k}{k_*}\right)^{n_t + \cdots} \; ,
    \label{eq:Pscalar_powerlaw}
\end{align}
\end{subequations}
where $k_*$ is a chosen pivot scale, $A_s$ denotes the amplitude at $k_*$, $n_s \equiv 1 + {\rm d}\ln {\cal P}_{\zeta}^{\rm inf}/{\rm d}\ln k$ refers to the spectral index, $r$ represents the tensor-to-scalar ratio, and $n_t$ is the tensor tilt. 

The observational results from Planck~\cite{Planck:2018jri}, ACT~\cite{ACT:2025fju}, and BICEP/Keck~\cite{BICEP:2021xfz} indicate that the scalar spectrum is nearly scale-invariant with a mild red tilt, and the tensor-to-scalar ratio is sufficiently close to zero. This supports the picture that the primordial fluctuations may be generated during a
slow-roll inflationary epoch. One can define the following parameters
\begin{align}
    \epsilon \equiv \frac{M^2_{\rm Pl}}{2}\left(\frac{V_\phi^{}}{V}\right)^2 \; , \quad \eta \equiv M^2_{\rm Pl} \frac{V^{}_{\phi\phi}}{V} \; , 
    \label{eq:slow-roll-para}
\end{align}
where $V_{\phi\phi} \equiv \partial^2V/\partial \phi^2$. Then the slow-roll conditions turn out to be $\epsilon \ll 1$ and $|\eta| \ll 1$. In this case, the inflationary observables can be derived as
\begin{align}
r \approx 16\epsilon_*\, ,\; 
n_s-1 \approx 2\eta_*- 6\epsilon_*  \, ,\;  A_s \approx \frac{1}{24\pi^2 M^4_{\rm Pl}}  \frac{V}{\epsilon^{}_*} \, ,
\label{eq:slow_roll_observ}
\end{align}
where $\epsilon_*$ and $\eta_*$ are calculated at $k_*$. Meanwhile, the number of e-folds $N_e \equiv \log(a_e/a_*)$  from the time when the $k_*$ mode exits the horizon to the end of inflation can be computed as
\begin{align}
    N_e = \frac{1}{M_{\rm Pl}^2}\int^{\phi_*}_{\phi_e} \frac{V}{V_\phi}{\rm d}\phi \; ,
    \label{eq:slow-roll-efolds}
\end{align}
where $\phi_e$ denotes the value of $\phi$ at which either $\epsilon$ or $|\eta|$ becomes greater than one. For the typical CMB pivot scale $k_* = 0.05~{\rm Mpc}^{-1}$, we have $N_e \simeq 50 - 60$, depending on the post-inflationary reheating history. Hereafter, we take $N_e = 55$ for definiteness unless specified otherwise.

In the {\it quadratic} hilltop inflation model with $p=2$, the slow-roll parameter $|\eta| = 2M^2_{\rm Pl}/\mu^2_{}$ is independent of $\phi$. The slow-roll condition $|\eta| \ll 1$ therefore requires $\mu \gtrsim M_{\rm Pl}$, implying that the inflaton field excursion is necessarily trans-Planckian. For $p>2$ with sub-Planckian excursion, the slow-roll inflation terminates when $|\eta| \simeq 1$, corresponding to a field value $\phi_e$ satisfying
\begin{align}
    p(p-1)\phi^{p-2}_e\simeq \mu^p/M^2_{\rm Pl} \; ,
    \label{eq:phi_end}
\end{align}
and the number of e-folds $N_e$ during inflation can be approximated as
\begin{align}
    N_e^{} \approx \frac{\mu^p}{p(p-2)M^2_{\rm Pl}} \phi^{2-p}_* \; .
    \label{eq:single-efolds}
\end{align}
For a given $N_e$, this yields
\begin{align}
    \epsilon_* \propto \left(\frac{1}{N_e}\right)^{\frac{2p-2}{p-2}}\left(\frac{\mu}{M_{\rm Pl}}\right)^{\frac{2p}{p-2}} \; ,
    \label{eq:single-r}
\end{align}
indicating that $r \approx 16\epsilon_*$ is negligibly small. Meanwhile,
\begin{align}
    n_s \approx 1 - \frac{2(p-1)}{p-2}\frac{1}{N_e} \; ,
    \label{eq:single-ns}
\end{align}
which leads to $n_s \approx 0.927$ and $0.945$ (assuming $N_e = 55$) for the \Cubic and \Quartic hilltop models, respectively, both of which lie outside the 95\% C.L. regions in the latest Planck~\cite{Planck:2018jri} and ACT~\cite{ACT:2025fju} results.

\subsection{The curvaton mechanism}
The above picture can change if we introduce a curvaton field $\sigma$~\cite{Linde:1996gt, Lyth:2001nq,  Moroi:2001ct, Moroi:2002rd, Lyth:2002my}. Since $m_\sigma \ll H$ during inflation, the curvaton is effectively frozen at $\sigma_{\rm inf}$. Its quantum fluctuations at the horizon exit are promoted to classical perturbations with a nearly flat spectrum. At this stage, however, its contribution to the total curvature perturbations is negligible because of its tiny energy density, $\rho^{}_\sigma  \simeq   m^2_\sigma \sigma^2_{\rm inf}/2$.\footnote{In the presence of the $\lambda^2 \phi^2 \sigma^4 /M^2_{\rm Pl}$ term, the curvaton field does not behave purely as non-relativistic matter. A sizable $\sigma^4$ term would delay the epoch at which $\rho_\sigma$ becomes dominant. In our numerical calculations, we have checked that at the end of inflation, when $\phi=\phi_e$, the curvaton energy density is indeed much smaller than the total energy density.}   After inflation ends, the Hubble parameter $H$ decreases as the Universe expands. When $H^2(t_{\rm osc}) \equiv H^2_{\rm osc} \simeq m^2_{\sigma}$, which gives $T_{\rm osc} \simeq \sqrt{m_{\sigma} M_{\rm Pl}}$, 
the curvaton starts oscillating around its minimum and behaves as non-relativistic matter with an isocurvature density perturbation. Using the condition $H^2_{\rm osc} \simeq m^2_{\sigma}$, we have $\rho_r (t_{\rm osc}) \simeq 3 m^2_{\sigma} M^2_{\rm Pl}$. 
Since $\sigma_{\rm inf}\ll M_{\rm Pl}$, we see that at the oscillation time $t_{\rm osc}$, the ratio of the curvaton energy density to the radiation energy density satisfies $(\rho_\sigma/\rho_r)|_{t_{\rm osc}}\simeq (\sigma_{\rm inf}/M_{\rm Pl})^2/6 \ll 1$. 

However, as $\rho_\sigma \propto a^{-3}$ while $\rho_{\rm r} \propto a^{-4}$, the curvaton energy density becomes significant after sufficiently many Hubble times, provided that the oscillations last long enough before the curvaton decays. The final curvature perturbations then depend on the ratio between the curvaton energy density and the radiation energy density at the curvaton decay time $t_{\rm dec}$ defined as
\begin{align}
\label{eq:R}
    R\equiv \left.\frac{\rho_\sigma}{\rho_{\rm r}}\right|_{t_{\rm dec}}  \simeq \frac{1}{6} \left(\frac{\sigma_{\rm inf}}{M_{\rm Pl}}\right)^2 \left(\frac{T_{\rm osc}  }{T_{\rm dec}}\right)  \; ,
\end{align}
where $T_{\rm dec}$ denotes the curvaton decay temperature, which depends on the decay rate of the curvaton $\Gamma_{\rm curv}$ through $\Gamma_{\rm curv}\simeq H(T_{\rm dec})$. We  can in turn get
\begin{align}
\label{eq:Tdecay-condition}
    \Gamma_{\rm curv}^2 \simeq \frac{(1+R)\rho_r(T_{\rm dec})}{3M^2_{\rm Pl}} \approx \frac{m_{\sigma}^2 \sigma_{\rm inf}^2}{6 M_{\rm Pl}^2}  \left(\frac{T_{\rm dec}}{T_{\rm osc}}\right)^3 \; ,
\end{align}
where in the second step we have assumed that $R\gg 1$ so that the curvaton gives rise to large modifications to primordial observables in our model. With the help of eq.~\eqref{eq:Tdecay-condition}, $T_{\rm dec}$ can be written as
\begin{align}
    T_{\rm dec}\simeq 2\sqrt{m_{\sigma} M_{\rm Pl}}  \, \left(\frac{M_{\rm Pl}}{\sigma_{\rm inf}}\right)^{\frac{2}{3}}\left(\frac{\Gamma_{\rm curv}}{m_{\sigma}}\right)^{\frac{2}{3}} \; .
\end{align}
Meanwhile, using eqs.~\eqref{eq:R} and~\eqref{eq:Tdecay-condition}, we can estimate the magnitude of $\Gamma_{\rm curv}$ in the large $R$ regime, namely,
\begin{align}
    \frac{\Gamma_{\rm curv}}{m_{\sigma}}\simeq 6\times 10^{-24} \, \left(\frac{\sigma_{\rm inf}}{10^{14}\,{\rm GeV}}\right)^4 \left(\frac{500}{R}\right)^{\frac{3}{2}}.
 \end{align}
Substituting this back into $T_{\rm dec}$, we obtain
\begin{align}
    T_{\rm dec}\simeq  1\, {\rm GeV}\left(\frac{\sigma_{\rm inf}}{10^{14}\,{\rm GeV}}\right)^2\left(\frac{500}{R}\right)\left(\frac{m_{\sigma}}{10^{6}\,{\rm GeV}}\right)^{\frac{1}{2}}\,.
\end{align}
Assuming an instant energy transfer from the curvaton to radiation, we obtain a temperature after the decay $T_{\rm RH} \sim R^{1/4}T_{\rm dec}$. Taking $\sigma_{\rm inf} \sim 10^{14}~{\rm GeV}$, $R \sim 500$, and $m_{\sigma} \sim 10^{6}~{\rm GeV}$, we obtain $T_{\rm RH} \sim  {\rm GeV}$, which is well above Big-Bang Nucleosynthesis (BBN) bound $T_{\rm RH} \gtrsim 4-5~{\rm MeV}$. Apparently, the determination of $R$ depends on $\Gamma_{\rm curv}$, which is rather model-dependent. In this work, instead of considering specific curvaton decay models, we regard $R$ as a free parameter for simplicity.

The total curvature perturbation $\zeta$ on super-horizon scales is conserved for purely adiabatic evolution. In our scenario, $\zeta_{\rm r}$ and $\zeta_\sigma$ denote the component curvature perturbations of the radiation and curvaton fluids, respectively. In the flat gauge, they are given by
\begin{subequations}
\begin{align}
    \zeta_{\rm r} & = -H \delta \rho_{\rm r}/\dot{\rho}_{\rm r} = \delta \rho_{\rm r}/(4\rho_{\rm r})\;, \\
    \zeta_\sigma & = -H \delta \rho_{\sigma}/\dot{\rho}_{\sigma} = \delta \rho_{\sigma}/(3\rho_{\sigma})\; ,
\end{align}
\end{subequations}
where $\zeta_\sigma$ is evaluated after the curvaton has entered its pressureless oscillating phase. Before the curvaton decay, $\zeta_{\rm r}$ is inherited from the inflaton decay products. At early times, when the curvaton energy density is negligible, one has $\zeta \simeq \zeta_{\rm r}$. As the curvaton energy fraction grows, however, $\zeta$ evolves due to the relative entropy perturbation between the two fluids. In the sudden-decay approximation, the total curvature perturbation after decay is therefore written as~\cite{Lyth:2001nq} 
\begin{align}
\label{eq:total-zeta}
    \zeta= \frac{4\zeta_{\rm r} + 3R\zeta_\sigma }{4+3R} = (1- r_{\rm dec})\zeta_{\rm r} + r_{\rm dec}\zeta_{\sigma}\; ,
\end{align}
where $r_{\rm dec} \equiv 3R/(4+3R)$. Unlike the standard curvaton mechanism, what we consider here is a mixed inflaton--curvaton scenario in which $\zeta_{\rm r}$ can not be neglected~\cite{Langlois:2004nn,Lazarides:2004we,Ichikawa:2008ne}. It is convenient to rewrite eq.~\eqref{eq:total-zeta} as
\begin{align}
\zeta = \zeta_{\rm r} + r_{\rm dec}\left(\zeta_{\sigma}-\zeta_{\rm r}\right),
\end{align}
so that the adiabatic mode induced by the inflaton is carried by $\zeta_{\rm r}$, while the curvaton provides an additional contribution through the relative isocurvature perturbation $\zeta_{\sigma}-\zeta_{\rm r}$. 
Accordingly, we define $\zeta_{\rm inf}\equiv \zeta_{\rm r}$ and $\zeta_{\rm curv}\equiv r_{\rm dec}\left(\zeta_{\sigma}-\zeta_{\rm r}\right)$. Then the curvature perturbation power spectrum can be expressed as 
\begin{align}
\label{eq:PS_def}
    \P_\zeta &=\P_\zeta^{\rm inf}+\P_{\rm \zeta}^{\rm curv}+\P_{\zeta}^{\rm mix}\notag\\
    &\propto \langle \zeta_{\rm inf}\zeta_{\rm inf} \rangle +\langle \zeta_{\rm curv}\zeta_{\rm curv} \rangle + 2\langle \zeta_{\inf }\zeta_{\rm curv} \rangle \; ,
\end{align}
where $\langle \cdots \rangle$ denotes the correlation function.\footnote{By definition, $\langle \zeta_{\bf k}\zeta^*_{\bf k^\prime} \rangle = (2\pi)^3 \delta^{(3)}({\bf k} - {\bf k}^\prime) P_{\zeta}(k)$. After horizon exit, each mode freezes and becomes effectively classical, so the power spectrum is given by the squared amplitude of the mode function at horizon exit $P_\zeta(k) = |\zeta_k^{\rm exit}|^2$. Then the dimensionless power spectrum is given by ${\cal P}_\zeta(k) = k^3 P_\zeta(k)/(2\pi^2)$.} 
In the flat gauge, the last term in eq.~\eqref{eq:PS_def} is proportional to $\langle \delta\phi \delta\sigma\rangle$. Since the mass mixing between $\delta\phi$ and $\delta\sigma$ is proportional to $\lambda^2$, $\langle \zeta_{\rm inf}\zeta_{\rm curv}\rangle$ should be negligibly small compared with the diagonal terms. Therefore, we can ignore the cross term,\footnote{A more explicit form of the cross term is given in appendix~\ref{app:deltaN}.} which results in $\P_\zeta \approx \P_\zeta^{\rm inf}+\P_\zeta^{\rm curv}$ with
\begin{subequations}
\begin{align}
\label{eq:inf-Pzeta}
     &\P_\zeta^{\rm inf}    = \left(\frac{H_*}{2\pi}\right)^2\frac{1}{2\epsilon_* M^2_{\rm Pl} }\; , \\
     \label{eq:curv-Pzeta}
     &\P_\zeta^{\rm curv}  =\left(\frac{R}{4+3R}\right)^2 \left(\frac{H_*}{\pi \sigma_{\rm inf}}\right)^2 \; .
\end{align}
\end{subequations}

Evaluating the above quantities at the pivot CMB scale $k_* $ gives the amplitudes
\begin{align}
    A_{s}^{\rm inf} \equiv \P_\zeta^{\rm inf} (k_*)\; , \qquad A_{s}^{\rm curv} \equiv \P_\zeta^{\rm curv}(k_*)\; .
\end{align}
The total amplitude is $A_s = A_{s}^{\rm inf}+A_{s}^{\rm curv}$. 
Above, we see that the total curvature perturbations have different contributions from the inflaton and the curvaton, both depending on the ratio $R$. 

With the help of eqs.~\eqref{eq:inf-Pzeta} and~\eqref{eq:curv-Pzeta}, we can derive the spectral index $n_s$ as
\begin{align}
   \label{eq:tot_ns}
    n_s - 1 &=  \left.\frac{1}{{\cal P}_{\zeta}}\left( \frac{{\rm d}{\cal P}_{\zeta}^{\rm inf} }{{\rm d}\ln k} + \frac{{\rm d}{\cal P}_{\zeta}^{\rm curv} }{{\rm d}\ln k}\right)\right|_{k_*} \\ \nonumber
    &= \frac{{A}_{s}^{\rm inf}}{ {A}_{{s}}}(n_{s}^{\rm inf} - 1) + \frac{{A}_{s}^{\rm curv}}{{A}_{{s}}}(n_{s}^{\rm curv} - 1) \; ,
\end{align}
where $n_{s}^{\rm inf}$ can be calculated using eq.~\eqref{eq:slow_roll_observ} in the slow-roll regime, whereas $n_{s}^{\rm curv}$ is given by~\cite{Lyth:2001nq}
\begin{align}
n_{s}^{\rm curv}-1 \approx  2 \frac{\dot{H}_*}{H_*^2} + \frac{2}{3} \frac{{m^2_{\sigma}}}{H_*^2} \; .
\end{align}
In the case where ${m_{\sigma}} \ll H_*$ and $\dot{H}_*/H^2_* \ll 1$, we approximately have $n_{s}^{\rm curv} \approx 1$. We can now readily see how the presence of the curvaton modifies the inflationary predictions of hilltop models. In the single-field hilltop scenario, the amplitude of the primordial scalar spectrum $A_s^{\rm inf}$ can be compatible with observations, while the predicted spectral index $n_s$ typically comes out too small in the sub-Planckian case. With the curvaton, both $A_s$ and $n_s$ are altered. The total amplitude $A_s$ no longer needs to be entirely sourced by $A_s^{\rm inf}$. Accordingly, one has $A_s^{\rm inf}/A_s<1$ in eq.~\eqref{eq:tot_ns}, which implies $|n_s-1|< |n_s^{\rm inf}-1|$, so that the predicted $n_s$ can be closer to unity compared to the single-field hilltop case. Using the best-fit values $n_s = 0.9649$ and $\ln(10^{-10}A_s) = 3.049$ from \texttt{Planck18} (TT,TE,EE+lowE) results for illustration, one arrives at $\Ainf \simeq 1.0 \times 10^{-9}$ and $\Acurv \simeq 1.1 \times 10^{-9}$ for the \Cubic hilltop, and $\Ainf \simeq 1.3 \times 10^{-9}$ and $\Acurv \simeq 7.6 \times 10^{-10}$ for the \Quartic hilltop.

In addition, the tensor-to-scalar ratio $r$ also gets modified, i.e.,
\begin{align}
     r = \frac{{A}_{t}}{{A}_{s}} = \frac{16\epsilon_*}{1 + {A}_{s}^{\rm curv}/{A}_s^{\rm inf}}    \; ,
    \label{eq:ten-to-sca}
\end{align}
where ${A}_t = 2H_*^2/(\pi^2 M^2_{\rm Pl})$ has been adopted.

In appendix~\ref{app:deltaN}, we re-derive the above results adopting the $\delta N$ formalism~\cite{Sasaki:1995aw,Lyth:2005fi}. Moreover, the $\delta N$ formalism could also help us calculate the magnitude of the primordial non-Gaussianity~\cite{Komatsu:2001rj,Bartolo:2004if,Lyth:2005fi,Chen:2010xka}, which may be significant in the curvaton scenario, since the final curvature perturbations arise from the nonlinear conversion of the curvaton isocurvature fluctuations into adiabatic ones after its decay. The local non-Gaussianity can be described by a dimensionless parameter $f_{\rm NL}$, which can be estimated as [cf. eq.~\eqref{eq:non-gause-expr}]
\begin{align}
    f_{\rm NL} \approx \left( \frac{\Acurv}{A_s}\right)^2 \left(\frac{5}{4r_{\rm dec}}-\frac{5}{3}-\frac{5}{6}r_{\rm dec}\right) \; .
    \label{eq:fnl_exp}
\end{align}

\section{Results and Discussions}
\label{sec:numerical-results}

In this section, we test the curvaton-assisted \Cubic and \Quartic hilltop inflation models against the CMB observations. Before proceeding to the numerical analysis, we shall make some analytical order-of-magnitude estimates.

We focus on the scenario where the primary inflationary phase can be approximately described by a purely \Cubic or \Quartic hilltop potential, for which the formulae for the primordial observables remain valid. To achieve this, we require $B_p \lesssim m_\phi^2$, so that $V_\phi$ is dominated by the second term in  eq.~\eqref{eq:V_phi}. 
By adopting the formulae in single-field slow-roll hilltop inflation, we can rewrite eqs.~\eqref{eq:inf-Pzeta} and~\eqref{eq:curv-Pzeta} as the expressions that depend explicitly on the model parameters. For $\Ainf$, we have
\begin{subequations}
\begin{align}
\label{eq:cubic-As}
    \Ainf \approx&  \frac{3N^4_e M^2_{\rm Pl}\Lambda^4}{4\pi^2 \mu^6} & (cubic~{\rm hilltop} ) \; , \\
    \label{eq:quartic-As}
    \Ainf \approx &  \frac{8N^3_e \Lambda^4}{3\pi^2 \mu^4}  & (quartic~{\rm hilltop} )\; ,
\end{align}
\end{subequations}
and
\begin{align}
    \Acurv \approx  \frac{\lambda \Lambda^4}{27\pi^2 m_\phi M^3_{\rm Pl}} \; .
    \label{eq:As-curv}
\end{align}
where we have assumed $R \gg 1$ and omitted the ${\cal O}(1)$ factor $\xi$. It can be seen that $\Lambda$ exhibits a positive relation with $\mu$, whereas it shows a negative relation with $\lambda$.  The ratio between $\Ainf$ and $\Acurv$ turns out to be
\begin{align}
    \frac{\Ainf}{\Acurv} =  \frac{81N^4_e m_\phi M^5_{\rm Pl} }{ 4\lambda  \mu^6}  \quad \text{and}
     \quad \frac{72N^3_e m_\phi M^3_{\rm Pl} }{ \lambda \mu^4} \; ,
     \label{eq:A_ratio}
\end{align}
for the \Cubic and \Quartic hilltop models, respectively. $\Lambda$ approximately cancels out in the ratio. One can thus see from eqs.~\eqref{eq:tot_ns} and \eqref{eq:ten-to-sca} that the dependence of $n_s$ and $r$ on $\Lambda$ is weak. In addition, we should require that $|V_{\sigma\sigma}| = m_\sigma^2 + 6\lambda^2\phi^2\sigma^2_{\rm inf}/M_{\rm Pl}^2\lesssim H^2
$ holds at $\phi_e$, which turns out to be a strong constraint on the parameter space. More accurately, one can see from eq.~\eqref{eq:phi_end} that $\phi^2_e \simeq \mu^6/(36M_{\rm Pl}^4) $ in the \Cubic case and $\phi^2_e \simeq \mu^4/(12M_{\rm Pl}^2) $ in the \Quartic case. After substituting the analytical estimate of $\phi_e$ into eq.~\eqref{eq:A_ratio}, we find that the \Cubic case carries one extra power of $N_e$ in $\Ainf/\Acurv$, compared with the \Quartic case. As a result, demanding $\Ainf$ and $\Acurv$ to be of comparable size typically pushes the \Cubic model towards a larger $\phi_e$, which makes the condition $|V_{\sigma\sigma}|\lesssim H^2$ much harder to satisfy.  

Based on the above discussions, we perform an illustrative Bayesian analysis to assess the consistency of our curvaton-assisted hilltop inflation model with cosmological observations. We fix the values of $m_\phi =  3\times 10^{-11} M_{\rm Pl}$, $m_\sigma =  5\times10^{-12} M_{\rm Pl}$, and $R = 500$,\footnote{
The choice of fixing the values of $\{ R, m_\phi, m_\sigma \}$ is motivated as follows. First, from eq.~\eqref{eq:curv-Pzeta}, a sizable curvaton contribution to the scalar power spectrum requires a sufficiently large $R$, while for $R\gg 1$ the prefactor $R/(4+3R)$ is already close to saturation and depends only weakly on the precise value of $R$. Second, $m_\phi$ affects $A_s^{\rm curv}$ mainly through the value of $\sigma_{\rm inf}\simeq \xi \sigma_{\rm t}$ at the exit from the diffusion region. On the one hand, according to eq.~\eqref{eq:curv-Pzeta}, obtaining a sizable $A_s^{\rm curv}$ disfavors very large $m_\phi$. On the other hand, as shown in appendix~\ref{app:cubic}, once $m_\phi^2\ll B_3$, $\sigma_{\rm t}$ becomes nearly independent of $m_\phi$. We therefore fix $m_\phi$ to a representative value. Finally, the role of $m_\sigma$ is mainly to ensure that the curvaton remains effectively frozen during hilltop inflation, and its impact on the inflationary observables is negligible within the parameter region of interest. } while treating the remaining parameters $\theta = \{ \mu, \Lambda, \lambda \}$ as the scanning parameters. A log-flat (uniform-in-log) prior is imposed on these parameters, such that $\pi(\theta)=1$ for
\begin{align}
    -1\leq   \log_{10}(\mu/M_{\rm Pl})   \leq 0 \; , &\quad
    -5\leq \log_{10}(\Lambda/M_{\rm Pl})  \leq -2 \; , \nonumber \\
     -5\leq \log_{10}\lambda   \leq -1 \; , &\quad
       (|V_{\sigma\sigma}|/H^2) |_{\phi_e}  < 1 \; .  
\end{align}
Otherwise, $\pi(\theta)=0$. Here we restrict the prior range to $\mu \lesssim M_{\rm Pl}$ to maintain a clean separation between the sub-Planckian and trans-Planckian regimes. However, we note that this is mainly a choice for definiteness. In hilltop inflation, even when $\mu$ is slightly larger than $M_{\rm Pl}$, e.g. $\mu \simeq 2 M_{\rm Pl}$, the actual inflaton excursion may still remain below the Planck scale.

\begin{figure}[t!]
        \centering
        \includegraphics[width=1.0\linewidth]{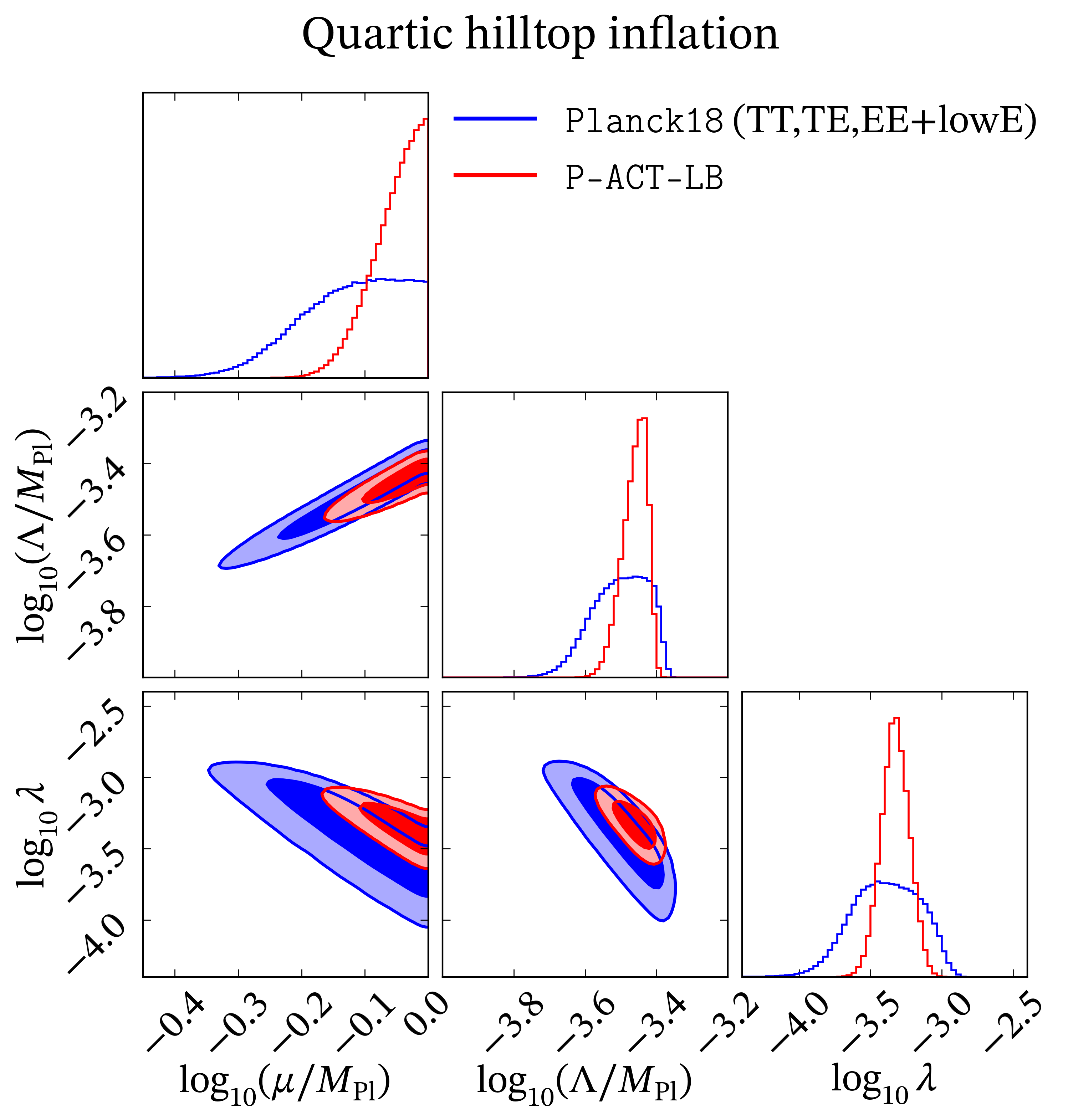}
        \caption{The 1D marginalized and 2D joint posterior probability distributions for the model parameters $\{\mu, \Lambda, \lambda\}$ in the \Quartic hilltop model, where we fix $R = 500$, $m_\phi =  3\times10^{-11} M_{\rm Pl}$, and $m_\sigma =  5 \times 10^{-12} M_{\rm Pl}$. Constraints are derived from an MCMC Bayesian analysis incorporating observational data for $n_s$ and ${\cal A}_s$ from \texttt{Planck18} (TT,TE,EE+lowE) (blue) and \texttt{P-ACT-LB} (red), as well as the upper limit on the tensor-to-scalar ratio $r < 0.036$ at 95\% C.L. from \texttt{BICEP/Keck18} (\texttt{BK18}). The inner and outer contours in the 2D plots correspond to the 68\%  and 95\% credible intervals, respectively.}
        \label{fig:quar_results}
\end{figure}

\begin{table}[t]
\centering
\caption{Marginalized posterior constraints for the curvaton assisted \Quartic hilltop model in the sub-Planckian region, where we set $R=500$, $m_\phi = 3\times 10^{-11}M_{\rm Pl}$ and $m_\sigma = 5 \times 10^{-12}M_{\rm Pl}$. 
Unless otherwise stated, quoted uncertainties correspond to the marginalized
68\% credible intervals. For $\mu$ in the \texttt{P-ACT-LB} case, we quote the 95\%
lower bound, since its posterior is visibly skewed toward the prior boundary
$\mu<M_{\rm Pl}$. For $r$, we quote the 95\% upper bound.}
\renewcommand{\arraystretch}{1.5}
\begin{tabular}{|c|c|c|}
\hline
\multirow{2}{*}{Parameters}
& \multicolumn{2}{c|}{\textit{Quartic} hilltop} \\
\cline{2-3}
& \texttt{Planck18} & \texttt{P-ACT-LB} \\
\hline

$\log_{10}(\mu/M_{\rm Pl})$ & $-0.113^{+0.077}_{-0.093}$ & $>-0.129$ \\
$\mu/M_{\rm Pl}$            & $0.771^{+0.149}_{-0.149}$ & $>0.743$ \\

$\log_{10}(\Lambda/M_{\rm Pl})$
& $-3.497^{+0.075}_{-0.086}$
& $-3.457^{+0.029}_{-0.041}$ \\

$\Lambda/M_{\rm Pl}$
& $(3.18^{+0.60}_{-0.57}) \times 10^{-4}$
& $(3.49^{+0.24}_{-0.32}) \times 10^{-4}$ \\

$\log_{10}\lambda$
& $-3.386^{+0.244}_{-0.243}$
& $-3.326^{+0.099}_{-0.098}$ \\

$\lambda$
& $(4.11^{+3.09}_{-1.76}) \times 10^{-4}$
& $(4.72^{+1.21}_{-0.95}) \times 10^{-4}$ \\

\hline
$n_s$
& $0.9630^{+0.0041}_{-0.0043}$
& $0.9720^{+0.0025}_{-0.0029}$ \\

${\cal A}_s$
& $3.045^{+0.016}_{-0.016}$
& $3.060^{+0.011}_{-0.011}$ \\

$r$
& $<10^{-6}$
& $<10^{-6}$ \\

$f_{\rm NL}$
& $-0.114^{+0.053}_{-0.066}$
& $-0.278^{+0.060}_{-0.058}$ \\
\hline
\end{tabular}
\label{tab:bestfit}
\end{table}

For the purpose of a phenomenological estimate, we construct a simplified likelihood in terms of the observables $D = \{\A_s, n_s, r \}$ (with $\A_s \equiv \ln(10^{10}A_s)$). For simplicity, we assume the measurement errors are Gaussian and uncorrelated, and then the log-likelihood is related to the $\chi^2$ statistic as $\log\mathcal{L} = -\chi^2/2$, where the total $\chi^2$  is calculated as the sum of the contributions from each observable
\begin{align}
\chi^2(\theta) = \frac{(n_s(\theta) - \bar{n}_s)^2}{\sigma_{n_s}^2} + \frac{(\A_s(\theta) - \bar{\A}_s)^2}{\sigma_{\A_s}^2} + \frac{r^2(\theta)}{\sigma_{r}^2} \; ,
\end{align}
where $n_s(\theta)$, $\A_s(\theta)$, and $r(\theta)$ are the values predicted by the model for the parameter set $\theta$, the values $\bar{n}_s$ and $\bar{\cal A}_s$ are the observed central values, and $\sigma$ represents the corresponding 1$\sigma$ (68\% C.L.) uncertainties. This simplified likelihood is sufficient for identifying  viable benchmark regions, but should not be regarded as a substitute for a full experimental likelihood analysis. As discussed above, we adopt the \texttt{Planck18} (TT,TE,EE+lowE) constraints and the combination of ACT DR6, Planck, and DESI Year-1 data, labeled by \texttt{P-ACT-LB} as the observational baselines for our parameter inference. For \texttt{Planck18} (TT,TE,EE+lowE), the central values and the corresponding 68\% C.L. uncertainties for $n_s$ and $\A_s$ at the pivot $k_* = 0.05~{\rm Mpc}^{-1}$ are determined by~\cite{Planck:2018jri} 
\begin{align}
    n_s = 0.9649 \pm 0.0044 \; , \quad \A_s = 3.045 \pm 0.016 \; ,
\end{align}
while for \texttt{P-ACT-LB}, we use\footnote{We approximate $\sigma_{\A_s}$ by the average value $\sigma_{\A_s} = 0.0115$ when constructing the likelihood function for $\A_s$.}~\cite{ACT:2025fju}
\begin{align}
    n_s = 0.9743 \pm 0.0034 \; , \quad \A_s = 3.060^{+0.011}_{-0.012} \; .
\end{align}
For the tensor-to-scalar ratio $r$, the most stringent upper bound is derived from the \texttt{BICEP/Keck18} (\texttt{BK18}) result~\cite{BICEP:2021xfz}, namely, 
\begin{align}
    r < 0.036~{\rm (95\%~C.L.)} \quad {\rm at}~k_* = 0.05~{\rm Mpc}^{-1} \; .
\end{align}
We thereby model the $\chi^2$ function for $r$ as a one-sided Gaussian distribution with a mean of zero and a standard deviation $\sigma_{r} = 0.036/1.96 = 0.0184$.

The posterior probability distribution $P(\theta|D)$ of the parameters $\theta$ given the observational data $D$ is proportional to the product of the likelihood function $\mathcal{L}(D|\theta)$ and the prior probability distribution $\pi(\theta)$
\begin{align}
    P(\theta|D) \propto \mathcal{L}(D|\theta) \pi(\theta) \; .
\end{align}
We explore the parameter space using an MCMC algorithm. For the \Cubic hilltop model, we do not find any viable parameter region with $\mu < M_{\rm Pl}$ in our numerical scans, even after varying $m_\phi$, $m_\sigma$ and $R$. The main issue is that the model cannot simultaneously reproduce the observed inflationary observables and satisfy the curvaton condition. A detailed analysis of the viability of the \Cubic hilltop model can be found in appendix~\ref{app:cubic}. In contrast, in the \Quartic model we find a parameter region that is consistent with current cosmological observations. The resulting 1D and 2D marginalized posterior distributions are displayed as corner plots in  Fig.~\ref{fig:quar_results}. The inner and outer contours correspond to the 68\% and 95\% credible intervals (i.e., approximately 68\% and 95\% of the samples fall within the corresponding contours), while the blue and red shaded regions represent the constraints from the \texttt{Planck18} and \texttt{P-ACT-LB}, respectively. Meanwhile, we also present marginalized posterior constraints for model parameters \{$\mu$, $\Lambda$, $\lambda$\} and the predicted values of $\A_s$, $n_s$, $r$ and $f_{\rm NL}$ in Table~\ref{tab:bestfit}. Some remarks are as follows.

Firstly, our analysis demonstrates that the curvaton-assisted \Quartic hilltop inflation model is consistent with current observations from both Planck and ACT collaborations.  It is evident that \texttt{P-ACT-LB} yields more compact posterior distributions, providing more stringent constraints on the model parameters.

Secondly, the preferred value of $\Lambda$ from the posterior distributions is around $ {\cal O}(10^{15})~\text{GeV}$,  implying a potential origin from Grand Unified Theories (GUTs).  The median value of the coupling parameter $\lambda$ is determined to be $\sim4 \times10^{-4}$. As for the parameter $\mu$, the 1D posterior is skewed toward the upper prior boundary $\mu < M_{\rm Pl}$. In the \texttt{Planck18} case, the distribution already shows a clear flattening before the imposed cutoff $\mu<M_{\rm Pl}$ is reached, and still admits a meaningful 68\% credible interval,
$\mu/M_{\rm Pl}=0.771^{+0.149}_{-0.149}$.
In contrast, the \texttt{P-ACT-LB} posterior is more strongly piled up against the boundary, with about 32\% of the samples satisfying $\mu/M_{\rm Pl}>0.95$, indicating a substantial prior-boundary effect. For this reason, we quote the constraint on $\mu$ in terms of a one-sided 95\% lower bound.

Finally, the parameter relations exhibited in the 2D posterior distributions indicate that $\mu$ is negatively correlated with $\lambda$ and positively correlated with $\Lambda$, which are well described by the analytical formulae in eqs.~\eqref{eq:cubic-As},~\eqref{eq:quartic-As} and~\eqref{eq:As-curv}. 

So far, our analysis has focused on the sub-Planckian hilltop regime. In that case, the inclusion of the curvaton alleviates the tension with observations, while the predicted tensor-to-scalar ratio remains negligible, $r<10^{-6}$. It is nevertheless instructive to briefly examine the trans-Planckian regime with $\mu \gtrsim M_{\rm Pl}$, where the inflationary dynamics is qualitatively different. In this case, the end of slow-roll inflation is no longer determined by $\eta=-1$, but instead by $\epsilon=1$. Accordingly, the field value at the end of inflation, $\phi_e$, is no longer given by eq.~\eqref{eq:phi_end}, and the corresponding $r-n_s$ relation is modified. For the hilltop potential $V(\phi)\propto 1 -(\phi/\mu)^p + \cdots$, one obtains at first order in the slow-roll approximation~\cite{Planck:2015sxf}
\begin{subequations}
\begin{align}
\label{eq:trans-r}
r &\approx \frac{8p^2 (M_{\rm Pl}/\mu)^2 x^{2p-2}}{(1-x^p)^2} \; , \\
\label{eq:trans-ns}
n_s -1 &\approx - \frac{2p(p-1)(M_{\rm Pl}/\mu)^2 x^{p-2}}{1-x^p} - \frac{3r}{8} \; ,
\end{align} 
\end{subequations}
where $x\equiv\phi_*/\mu$. In the limit $\mu \gg M_{\rm Pl}$, this reduces to $n_s \approx 1 - 3r/8$.

\begin{figure}[t!]
        \centering
        \includegraphics[width=1\linewidth]{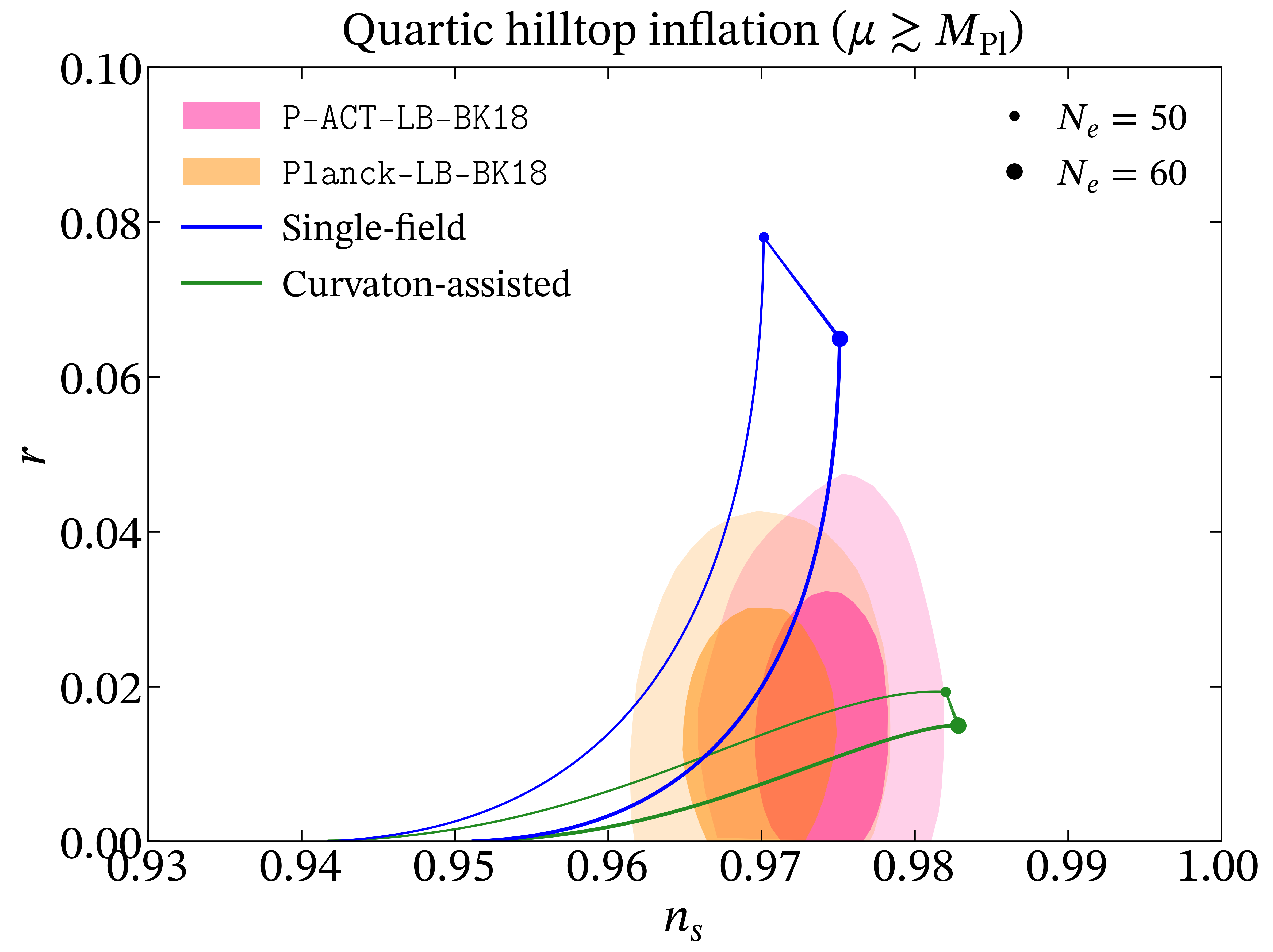}
        \caption{Relations between the tensor-to-scalar ratio $r$ and the spectral index $n_s$ in the trans-Planckian \Quartic hilltop model where $\mu \gtrsim M_{\rm Pl}$. The pink- and orange-shaded contours are the constraints from the \texttt{P-ACT-LB-BK18} and \texttt{Planck-LB-BK18} results, with the darker and lighter colors denoting the 68\% and 95\% C.L. allowed regions, respectively~\cite{AtacamaCosmologyTelescope:2025nti}. Blue lines correspond to the predicted $r-n_s$ relations in the single-field hilltop models, while the green lines represent the  $r - n_s$ relations in the curvaton-assisted \Quartic hilltop model for $\lambda=10^{-6}$, obtained from the viable points with ${\cal A}_s$ restricted to the narrow range $3.0<{\cal A}_s<3.1$, by averaging $r$ in each $n_s$ bin. Note that here the results for the curvaton-assisted hilltop model are also derived within the trans-Planckian regime.}
        \label{fig:ns-r-relation}
\end{figure}

The blue curves in Fig.~\ref{fig:ns-r-relation} show the corresponding $r$--$n_s$ relations for the single-field \Quartic hilltop inflation model, together with the experimental constraints from \texttt{P-ACT-LB-BK18} and \texttt{Planck-LB-BK18}. In this regime, $r$ is significantly enhanced, leading to stronger tension with the data, especially once ACT data are included. Even for the $N_e=60$ case, the theoretical prediction lies almost entirely outside the 68\% credible region of the \texttt{P-ACT-LB-BK18} constraints.

For comparison, we also illustrate the behavior of the curvaton-assisted model in the trans-Planckian regime. This part is intended only as an illustrative extension, rather than as a full Bayesian analysis analogous to the sub-Planckian case discussed above. As a representative example, we fix $\lambda=10^{-6}$ and select points satisfying the basic observational requirement $3.0<{\cal A}_s<3.1$. The green curves in Fig.~\ref{fig:ns-r-relation} are then obtained by averaging $r$ within each $n_s$ bin for these selected points. One sees that $r$ is suppressed relative to the single-field case, giving typically $r\sim \mathcal{O}(0.01)$ for $n_s\sim 0.97$. This suggests that, unlike the single-field trans-Planckian hilltop model, the curvaton-assisted scenario can remain compatible with current constraints while predicting a potentially observable tensor signal, making it an interesting target for future CMB experiments such as LiteBIRD~\cite{LiteBIRD:2022cnt} and CMB-S4~\cite{Chang:2022tzj}.

Before closing this section, we turn to a brief comment on the  non-Gaussianity. According to Table~\ref{tab:bestfit}, the preferred local non-Gaussianity coefficient $f_{\rm NL}$  is found to be negative with a magnitude of ${\cal O}(0.1)$. The up-to-date constraint on the local non-Gaussianity is $f_{\rm NL} = -0.9 \pm 5.1$~\cite{Planck:2019kim}. Hence our prediction for the non-Gaussianity lies well within current experimental limits, yet still represents a small but distinct deviation from the standard single-field slow-roll prediction $f_{\rm NL} \sim {\cal O}(1-n_s)$.

\section{Conclusions}
\label{sec:con}

Conventional hilltop inflation models face the initial-condition problem that the inflaton field must start to roll extremely close to the top of the hilltop potential, raising a concern about its naturalness. In this work, we have discussed a mechanism involving a curvaton field that significantly relaxes this problem. We have focused on the \Cubic and \Quartic hilltop models in particular. The curvaton $\sigma$ couples to the inflaton field via a cross term given in eq.~\eqref{eq:V_C}, which contributes to the effective mass of the inflaton and modulates the shape of the inflaton potential. The required operators can be embedded in a supersymmetric framework.

Given white-noise fluctuations at the Hubble scale for both inflaton and curvaton fields, we have shown, both analytically and numerically, that the onset of hilltop inflation is insensitive to the initial position of the curvaton within a certain region of field space.
The dynamics of the field evolution is described by the Langevin equations. In the numerical simulation, we applied an MCMC analysis to solve the equations, and in the analytical derivation, mean-square values of the fields $\phi$ and $\sigma$ were taken as observables. Both results match very well, as shown in Fig.~\ref{fig:trajectories}, and the insensitivity to the initial conditions is confirmed in Fig.~\ref{fig:RMS-paths}. 

The curvaton is assumed to be sufficiently light compared with the Hubble rate during inflation, so that it does not dominate the Universe until the Hubble rate decreases to become comparable to the curvaton mass after inflation. It plays an essential role in modifying the primordial curvature perturbations via the ratio 
$R$ of the curvaton energy density to the radiation energy density at the time of curvaton decay. We have explicitly derived the expressions for the primordial observables in the presence of a curvaton. 

We further performed an illustrative Bayesian analysis, based on a simplified likelihood constructed from the Planck and recent ACT results, to confirm the consistency of our model with cosmological observations. Given a sufficiently large 
$R$ and suitable inflaton and curvaton masses, the overall potential scale 
$\Lambda$ can be of order the GUT scale, and the inflaton field value can be of order the sub-Planckian scale, both remaining consistent with these data in the \Quartic hilltop model. Furthermore, we compared our model with single-field hilltop inflation in terms of the predicted tensor-to-scalar ratio and spectral index in the trans-Planckian regime. A tensor-to-scalar ratio of ${\cal O}(0.01)$ was obtained in the curvaton-assisted inflation scenario with $\mu \gg M_{\rm Pl}$ and $\lambda = 10^{-6}$, consistent with both the Planck and the recent ACT data, and this can be tested in next-generation CMB experiments.

\begin{acknowledgements}
We are grateful to Stefan Antusch for useful discussions. XW thanks Changyuan Yao for valuable discussions on coding. WA is supported by
the European Union (ERC, NLO-DM, 101044443).
SFK acknowledges the STFC Consolidated Grant
ST/X000583/1 and thanks IFIC, Valencia, for hospitality;
his work was funded by a Leverhulme Trust Emeritus Fellowship Grant.
XW acknowledges the Royal Society as the funding source of the Newton International Fellowship; his work is partially funded by the European Union, NextGenerationEU, National Recovery and Resilience Plan (mission 4, component 2)
under the project {\it MODIPAC: Modular Invariance in Particle Physics and Cosmology} (CUP C93C24004940006). 
YLZ is partially supported by the National Natural Science Foundation of China (NSFC) under Grant Nos. 12535007, 12547104, and Zhejiang Provincial Natural Science Foundation of China under Grant No. LDQ24A050002. 
\end{acknowledgements}

\appendix
\section{Langevin Equations} \label{app:langevin}
In the stochastic inflation formalism, the evolution of the coarse-grained scalar field is governed by two contributions: a deterministic drift corresponding to its classical slow-roll motion, and a stochastic source arising from the continuous horizon crossing and subsequent freezing of short-wavelength quantum fluctuations. As a result, the dynamics of the field can be described by a stochastic differential equation of Langevin type~\cite{Starobinsky:1986fx}
\begin{align}
    {\rm d}\varphi = -\frac{V_\varphi}{3H^2} {\rm d}N + \frac{H}{2\pi} {\rm d}W_N \; ,
    \label{eq:langevin_general}
\end{align}
where $\varphi = \phi, \sigma$ and  $W_N$ denotes the Wiener stochastic process with increments satisfying $\langle {\rm d} W_N \rangle = 0$ and $\langle ({\rm d} W_N )^2\rangle = {\rm d} N$. ${\rm d} W_N$ is related to the white noise $\xi_\varphi$ via ${\rm d} W_N = \xi_\varphi {\rm d} N$. Then one can immediately note that eq.~\eqref{eq:langevin_general} takes the same form as  eqs.~\eqref{eq:langevin_phi} and \eqref{eq:langevin_sigma}.

Taking the expectation value of eq.~\eqref{eq:langevin_general} and utilizing $\langle {\rm d} W_N \rangle = 0$, we obtain the Langevin equation of the expectation value $\langle \varphi \rangle$ as
\begin{align}
    \frac{{\rm d}\evvphi}{{\rm d}N} = -\frac{\langle V_\varphi \rangle}{3H^2} \; .
    \label{eq:ev}
\end{align}

In order to derive the Langevin equation for the field mean-square values, we consider the stochastic differential of $\varphi^2$, namely,
\begin{align}
    \label{eq:seorder}
    {\rm d}(\varphi^2) = & 2\varphi \, {\rm d}\varphi + ({\rm d}\varphi)^2  \\
    = & -\varphi \frac{2 V_\varphi }{3H^2} {\rm d}N + 2\varphi \frac{H}{2\pi} {\rm d}W_N + \left(\frac{H}{2\pi}\right)^2   {\rm d}N  \nonumber \; ,
\end{align}
where $(\d \varphi )^2$ cannot be neglected since $\langle ({\rm d} W^{}_N )^2\rangle = {\rm d} N$ is of first order in the infinitesimal. Again, taking the expectation value of eq.~\eqref{eq:seorder} and using $\langle \varphi \, \d W_N \rangle = 0$, we obtain Langevin equation for $\vavphi$
\begin{align}
   \frac{\d \vavphi }{\d N} = - \frac{2}{3H^2}\left\langle   V_\varphi \cdot\varphi  \right\rangle + \left(\frac{H}{2\pi}\right)^2 \; . 
   \label{eq:va}
\end{align}

For our model setup with $\lambda \lesssim 10^{-2}$, the covariance between $\phi$ and $\sigma$ is negligible, which allows us to treat the Langevin equations for $\phi$ and $\sigma$ separately. As mentioned in the main text, the starting point of $\sigma$ is randomly selected in the range $\sigma \gtrsim \sigma_{\rm c}$, while the initial value of $\phi$ is very close to zero. In this case, $V^{}_\sigma \approx m^2_\sigma \sigma$ and eq.~\eqref{eq:va} reduces to
\begin{align}
    \frac{\d \vasigma }{\d N} \approx - \frac{2 m^2_\sigma}{3H^2}\langle  \sigma^2 \rangle - \frac{4\lambda^2 \langle\phi^2\rangle \langle \sigma^4 \rangle}{3 H^2 M_{\rm Pl}^2}   + \left(\frac{H}{2\pi}\right)^2 \; , 
   \label{eq:va-sigma}
\end{align}
which reproduces eq.~\eqref{eq:langevin_sigma_var_app}.
As for the field $\phi$, since $V_\phi$ contains higher-order terms in $\phi$, in general one cannot reduce the equation for $\vaphi$ into a form analogous to eq.~\eqref{eq:va-sigma}. In the following, we examine the explicit forms of Langevin equations for $\vaphi$ for the \Cubic and \Quartic hilltop cases.

{\bf Cubic hilltop.} In the \Cubic hilltop case with $p=3$, $V_\phi$ is written as
\begin{align}
   V_\phi = m^2_\phi \left(\frac{\sigma^4_{}}{\sigma^4_{\rm c}}-1\right) \phi - 3\Lambda^4_{}
   \frac{\phi^2_{}}{\mu^3_{}} + \cdots \; .
\end{align}
From eq.~\eqref{eq:ev}, it is not difficult to see that $\phi$ can develop a nonzero expectation value due to the $\phi^2$ term in $V_\phi$. A static solution can be reached if $\d\evphi/\d N \simeq 0$ is satisfied, which yields 
\begin{align}
    \evphi \simeq \frac{3\Lambda^4\vaphi}{m^2_\phi\mu^3}\left(\frac{\sigma^4_{}}{\sigma^4_{\rm c}}-1\right)^{-1} \; .
\end{align}
For the parameter range of interest, $\evphi \lesssim \phi_{\rm rms} \equiv \sqrt{\vaphi}$ holds. Hence, under the condition that $\phi$ follows an approximately Gaussian distribution, one can get $\langle \phi^3 \rangle \approx 3\evphi \cdot \vaphi$. Then eq.~\eqref{eq:va} can be recast into
\begin{align}
   \frac{\d \vaphi }{\d N} \approx - \frac{2}{3H^2}\left(  C_1 \vaphi + C_2 \vaphi^2  \right)+ \left(\frac{H}{2\pi}\right)^2 \; , 
   \label{eq:va-cubic}
\end{align}
with
\begin{align}
    C_1  = m^2_\phi \left(\frac{\langle\sigma^4_{}\rangle}{\sigma^4_{\rm c}}-1\right) \; , \quad C_2  = -\frac{27\Lambda^8}{m^2_\phi \mu^6_{}} \left(\frac{\langle\sigma^4_{}\rangle}{\sigma^4_{\rm c}}-1\right)^{-1} .
\end{align}

{\bf Quartic hilltop.} Unlike the \Cubic hilltop potential, in the \Quartic hilltop case we have
\begin{align}
   V_\phi = m^2_\phi \left(\frac{\sigma^4_{}}{\sigma^4_{\rm c}}-1\right) \phi - 4\Lambda^4_{}
   \frac{\phi^3_{}}{\mu^4_{}} + \cdots \; ,
\end{align}
which involves only odd powers of $\phi$. Therefore, as long as $\evphi$ is sufficiently close to zero, it will eventually stabilize at $\evphi = 0$. For an approximate Gaussian distribution, $\langle \phi^4 \rangle = 3\vaphi^2$ should be satisfied. In this case, the Langevin equation for $\vaphi$ also takes the form as
\begin{align}
   \frac{\d \vaphi }{\d N} \approx - \frac{2}{3H^2}\left(  C_1 \vaphi + C_2 \vaphi^2  \right)+ \left(\frac{H}{2\pi}\right)^2 \; , 
   \label{eq:va-quartic}
\end{align}
with
\begin{align}
    C_1  = m^2_\phi \left(\frac{\langle\sigma^4_{}\rangle}{\sigma^4_{\rm c}}-1\right) \; , \quad C_2  = -\frac{12\Lambda^4}{\mu^4_{}} \; .
\end{align}

\section{$\delta N$ formalism}\label{app:deltaN}
In this section, we derive the explicit formulae for inflationary observables in our model by implementing the $\delta N$ formalism~\cite{Sasaki:1995aw,Lyth:2004gb,Lyth:2005fi}. On super-horizon scales, the curvature perturbation $\zeta$ appears in the spatial part of the perturbed metric $\d s^2$ in the form
\begin{equation}
\d s^2 = -\d t^2 + a^2(t) e^{2\zeta(t,\vec{x})} \delta_{ij} \d x^i \d x^j \; .
\end{equation}
Accordingly, the physical volume element $\d V(\vec{x})$ around the position $\vec{x}$ is proportional to $a^3(t)e^{3\zeta(t,\vec{x})}$. On the other hand, the expansion of the background volume can be connected to the number of e-folds as $a^3(t) \sim e^{3N}$. As a result, $\d V(\vec{x}) \propto e^{3\left[N + \delta N(\vec{x})\right]}$ with $\delta N(\vec{x})$ being the difference between the local number of e-folds in each Hubble patch and the background average.
Hence one can directly identify the curvature perturbation with the local fluctuation in the number of e-folds, namely, $\zeta(\vec{x}) = \delta N(\vec{x})$.

In practice, one can select a spatially flat slice where the curvature perturbation in the three-dimensional metric vanishes as the initial hypersurface, labeled by ``*'', and a uniform-density slice where $\delta \rho = 0$ as the final hypersurface, labeled by ``$f$''. Note that the flat hypersurface must be evaluated at the horizon exit. Then we have $\zeta = \delta N^f_*$.  In the two-field framework, $\delta N^f_*$ can be computed with the help of Taylor expanding~\cite{Lyth:2004gb}
\begin{align}
    \delta N^f_* = N^{}_i\delta \varphi^i_* + \frac{1}{2}N^{}_{ij}\delta \varphi^i_* \delta \varphi^j_* 
    + \cdots\; ,
\end{align}
with $N_i \equiv \partial N^f_*/\partial \varphi^i_*$, $N_{ij} \equiv \partial^2 N^f_*/(\partial \varphi^i_* \partial \varphi^j_*) $, and $\varphi = \phi, \sigma$ in our case. We adopt the sudden–decay approximation for the curvaton and take the final uniform–density hypersurface to coincide with the epoch of its decay, and then $N_\phi$ and $N_\sigma$ can be evaluated by
\begin{align}
    N_\phi = \frac{1}{M_{\rm Pl}\sqrt{2\epsilon_*}} \; , \quad N_\sigma =  \frac{2r_{\rm dec}}{3\sigma_{\rm inf}} \; .
    \label{eq:N_i}
\end{align}
where
\begin{align}
    r_{\rm dec} \equiv \left.\frac{3\rho_\sigma}{4\rho_r + 3\rho_\sigma}\right|_{\rm decay} \; .
    \label{eq:r_dec_2}
\end{align}
The factor $r_{\rm dec}$ enters the expressions for $N_\phi$ and $N_\sigma$ because, after the curvaton decay, the total curvature perturbation $\zeta$ is a weighted combination of the adiabatic and isocurvature contributions, namely, $\zeta = \zeta_r + r_{\rm dec}(\zeta_\sigma-\zeta_r)$.

Now we can use the above formulae to calculate corresponding inflationary observables.

{\bf Power spectrum.}  The total power spectrum reads
\begin{align}
    {\cal P}_\zeta & \equiv \langle \zeta \zeta \rangle =  N_i N_j \langle \delta\varphi^i_* \delta\varphi^j_*\rangle \nonumber \\ 
    & = \left(\frac{H_*}{2\pi}\right)^2 (N^2_\phi + N^2_\sigma + 2\kappa N_\phi N_\sigma ) \; ,
    \label{eq:PS_deltaN}
\end{align}
where the coefficient $\kappa \equiv \langle \delta \phi_* \delta\sigma_{\rm inf} \rangle/[H^2_*/(2\pi)^2]$. Under the assumption that $\sigma$ is light and the mixing between $\phi$ and $\sigma$ is weak, $ \kappa \sim M^2_{\rm Pl} V_{\phi\sigma}/V = 4\lambda^2\phi \sigma^3/V \ll 1$ can be safely neglected. Hence we get
\begin{align}
    {\cal P}_\zeta \approx \left(\frac{H_*}{2\pi}\right)^2 \left[ \frac{1}{2\epsilon_* M^2_{\rm Pl}} + \left( \frac{2r_{\rm dec}}{3\sigma_{\rm inf}}\right)^2\right] \; ,
\end{align}
which turns out to be eqs.~\eqref{eq:inf-Pzeta} and \eqref{eq:curv-Pzeta}.

{\bf Tensor-to-scalar ratio.} As the introduction of the curvaton only modifies the scalar perturbation power spectrum, the tensor-to-scalar ratio becomes
\begin{align}
    r \equiv \frac{{\cal P}_{\cal T}}{{\cal P}_{\zeta}} = \frac{8}{M^2_{\rm Pl}(N^2_\phi + N^2_\sigma)} \; .
\end{align}
Substituting eq.~\eqref{eq:N_i} into the above equation, we obtain
\begin{align}
    r = \frac{16 \epsilon_*}{1 + 2\epsilon_* M^2_{\rm Pl}[2r_{\rm dec}/(3\sigma_{\rm inf})]^2} \; .
\end{align}

{\bf Spectral index.} In the $\delta N$ formalism, the spectral index can be expressed as
\begin{align}
n_s  =  1 & + \frac{N_\phi^2}{N_\phi^2+N_\sigma^2}\left(-6\epsilon_*+2\eta_*\right) \nonumber \\
& +\frac{N_\sigma^2}{N_\phi^2+N_\sigma^2}\left(-2\epsilon_*+2\eta_*^{(\sigma)}\right) \; ,
\end{align}
where $\eta^{(\sigma)} \equiv M^2_{\rm Pl}V_{\sigma\sigma}/V$ has been defined. Since the curvaton potential is rather flat, $\eta^{(\sigma)}$ should be highly suppressed. Moreover, $\epsilon_*$ should also be negligible in hilltop inflation models. Hence the expression for $n_s$ can be simplified into
\begin{align}
    n_s \approx 1 + 2\eta_*\frac{1}{1 + 8r^2_{\rm dec}\epsilon_* M^2_{\rm Pl}/(3\sigma_{\rm inf})^2} \; .
\end{align}

{\bf Non-Gaussianity.} If the primordial fluctuations obey a strict Gaussian distribution, all statistical information is fully characterized by the two-point function. To be more concrete, all odd-point correlation functions vanish, while even-point functions can always be reduced to products of two-point functions. In this sense, the presence of a non-vanishing three-point function serves as a clear indicator of non-Gaussianity. In the $\delta N$ formalism, a non-vanishing quadratic term may generate a nonzero three-point function. Schematically, one finds $\langle \zeta^3 \rangle \sim \langle \delta \varphi^i \delta \varphi^j \rangle \langle \delta \varphi^k \delta \varphi^l \rangle$, which implies that non-linear dependence of $\zeta$ on the field fluctuations could lead to non-Gaussianity~\cite{Komatsu:2001rj,Bartolo:2004if,Lyth:2005fi,Chen:2010xka}.

The non-Gaussianity can be described by a dimensionless parameter $f_{\rm NL}$, defined as
\begin{align}
    f_{\rm NL} & \equiv \frac{5}{6}\frac{N_i N_j N_{ij}}{(N_k N_k)^2} \nonumber \\ 
    & =  \frac{5}{6} \frac{N^2_\phi N_{\phi\phi} + 2N_\phi N_\sigma N_{\phi \sigma}+N^2_\sigma N_{\sigma\sigma}}{(N^2_\phi + N^2_\sigma)^2}\; .
    \label{eq:non-gauss}
\end{align}
In the case where $N_{\phi\phi}$ is small and $N_{\phi\sigma}$ is negligible, $f_{\rm NL}$ is reduced to
\begin{align}
    f_{\rm NL} \approx \frac{5}{6}\frac{N^2_\sigma N_{\sigma\sigma}}{(N^2_\phi + N^2_\sigma)^2} \; .
\end{align}

To proceed, we need the explicit form of $N_{\sigma\sigma}$. Keeping in mind that $r_{\rm dec}$ also depends on $\sigma_{\rm inf}$ through $\rho_\sigma \propto \sigma_{\rm inf}^2$, we have
\begin{equation}
N_{\sigma\sigma} = \frac{\partial}{\partial\sigma_{\rm inf}}
\left(\frac{2}{3}\frac{r_{\rm dec}}{\sigma_{\rm inf}}\right)
= \frac{2}{3}\left(\frac{r'_{\rm dec}}{\sigma_{\rm inf}}-\frac{r_{\rm dec}}{\sigma_{\rm inf}^2}\right) \; ,
\end{equation}
with $r'_{\rm dec}\equiv\partial r_{\rm dec}/\partial\sigma_{\rm inf}$. Using the definition
of $r_{\rm dec}$ given in eq.~\eqref{eq:r_dec_2}, one finds
\begin{equation}
N_{\sigma\sigma} = \frac{2r_{\rm dec}}{9\sigma_{\rm inf}^2}\,
\Big(3-4r_{\rm dec}-2r_{\rm dec}^2\Big)\, .
\end{equation}
Therefore, we eventually arrive at
\begin{align}
    f_{\rm NL} \approx \left( \frac{N^2_\sigma}{N^2_\phi + N^2_\sigma}\right)^2 \left(\frac{5}{4r_{\rm dec}}-\frac{5}{3}-\frac{5}{6}r_{\rm dec}\right) \; ,
    \label{eq:non-gause-expr}
\end{align}
where we should note that $\Acurv = H^2_* N^2_\sigma/(2\pi)^2$ and $\Ainf = H^2_* N^2_\phi/(2\pi)^2$.

\section{Viability of the cubic hilltop case}
\label{app:cubic}
In this appendix, we explain why the sub-Planckian \Cubic hilltop case cannot be rescued by the curvaton mechanism in our benchmark setup. As we have mentioned in the main text, the main difficulty arises from the curvaton condition $|V_{\sigma\sigma}| \lesssim H^2$. For the \Cubic hilltop model, using $\phi_e^2 \simeq \mu^6/(36M_{\rm Pl}^4)$ and $\sigma_{\rm inf}\simeq  \sigma_{\rm t}$ with
\begin{align}
\sigma_{\rm t}=\sigma_{\rm c}\left(1+\frac{B_3}{m_\phi^2}\right)^{1/4}\; ,\quad
\sigma_{\rm c}^2=\frac{m_\phi M_{\rm Pl}}{\lambda}\; ,
\end{align}
from $|V_{\sigma\sigma}| \lesssim H^2$ one finds
\begin{align}
\frac{\lambda m_\phi\mu^6}{6M_{\rm Pl}^5}
\left(1+\frac{B_3}{m_\phi^2}\right)^{1/2}
\lesssim \frac{\Lambda^4}{3M_{\rm Pl}^2} \; ,
\end{align}
where $m^2_\sigma$ has been omitted from $|V_{\sigma\sigma}|$. Hence $\lambda$ should obey the following upper bound
\begin{align}
\lambda \lesssim
\frac{2\Lambda^4 M^3_{\rm Pl}}
{m_\phi \mu^6 (1+B_3/m_\phi^2)^{1/2}} \; .
\end{align}
From Eq.~\eqref{eq:cubic-As}, we have
\begin{align}
A_s^{\rm inf}\approx \frac{3N_e^4M^2_{\rm Pl}\Lambda^4}{4\pi^2\mu^6} \; .
\end{align}
With the help of the above equation, the upper bound on $\lambda$ can be rewritten as
\begin{align}
\lambda \lesssim
\frac{8\pi^2 M_{\rm Pl}}{3N_e^4 m_\phi(1+B_3/m_\phi^2)^{1/2}}
A_s^{\rm inf}.
\label{eq:lambda-upper}
\end{align}
For our benchmark values $m_\phi =  3\times 10^{-11} M_{\rm Pl}$ and $m_\sigma =  5\times10^{-12} M_{\rm Pl}$, and requiring $A_s^{\rm inf}\sim 10^{-9}$, this gives
\begin{align}
\lambda \lesssim
\frac{10^{-4}}
{(1+B_3/m_\phi^2)^{1/2}}
\lesssim 10^{-4} \; .
\end{align}
One should note that although the upper bound in eq.~\eqref{eq:lambda-upper} appears to scale as $\lambda \propto m_\phi^{-1}$, this does not imply that one can evade the constraint by taking $m_\phi$ arbitrarily small. Indeed, once $m_\phi^2 \ll B_3$, one has $(1+B_3/m_\phi^2)^{1/2}\simeq (B_3)^{1/2}/m_\phi$. This inverse dependence on $m_\phi$ cancels the explicit factor of $m_\phi$ in $\sigma^2_{\rm c}$, so that $\sigma_{\rm t}$ becomes approximately independent of $m_\phi$. The bound on $\lambda$ then saturates to a value independent of $m_\phi$, showing that lowering $m_\phi$ does not help to restore a sizable curvaton contribution.

On the other hand, the ratio between the curvaton and inflaton contributions to the scalar spectrum is
\begin{align}
\frac{A_s^{\rm curv}}{A_s^{\rm inf}}
\approx
\frac{4\lambda\mu^6}{81N_e^4 m_\phi M^5_{\rm Pl}}\; .
\end{align}
Since $\mu<M_{\rm Pl}$ and $\lambda$ is bounded as above, one obtains the conservative upper bound $ A_s^{\rm curv}/A_s^{\rm inf} \lesssim 10^{-2}$. Therefore, in the sub-Planckian \Cubic hilltop case, the curvaton contribution can at most be at the few-percent level and is, in general, much smaller. The inflaton contribution thus dominates the total scalar spectrum, and the prediction for $n_s$ remains essentially that of the single-field \Cubic hilltop model. 

We have also performed a full numerical scan of the \Cubic case. The results show that the posterior distribution of $n_s$ is sharply concentrated around $n_s \simeq 0.93$, which is precisely the characteristic prediction of the single-field sub-Planckian \Cubic hilltop model for $N_*=55$. This confirms the validity of the above analytical estimate, and further demonstrates that the inclusion of the curvaton is insufficient to bring the sub-Planckian \Cubic hilltop model into agreement with observations.

For the trans-Planckian case of \Cubic hilltop inflation, the results are similar to those of \Quartic trans-Planckian hilltop inflation. In this regime, the end of inflation is no longer determined by the condition $|\eta|=1$, but instead by $\epsilon=1$. The inflationary observables $n_s$ and $r$ are then governed by eqs.~\eqref{eq:trans-r} and~\eqref{eq:trans-ns} with $p=3$. In the limit $\mu \gg M_{\rm Pl}$, the single-field \Cubic hilltop model in the trans-Planckian case can already be compatible with current observations, and it typically predicts a sizable tensor-to-scalar ratio $r$.
Once the curvaton is included, 
for the trans-Planckian case of \Cubic hilltop inflation,
the value of $r$ can be suppressed, so that both $n_s$ and $r$ can simultaneously lie within the 68\% credible regions preferred by the \texttt{P-ACT-LB-BK18} and \texttt{Planck-LB-BK18} data.
In summary, the resulting 
$n_s-r$ predictions for the
trans-Planckian case of \Cubic hilltop inflation
are qualitatively similar to those shown in Fig.~\ref{fig:ns-r-relation} for the \Quartic hilltop trans-Planckian case. Thus, we do not display the results for the trans-Planckian case of \Cubic hilltop inflation separately since the plot would be very similar to Fig.~\ref{fig:ns-r-relation} for the \Quartic hilltop trans-Planckian case.

\bibliography{ref}

\end{document}